\documentclass[a4paper,fleqn,usenatbib]{mnras}

\usepackage{newtxtext,newtxmath}

\usepackage[T1]{fontenc}
\usepackage{ae,aecompl}

\usepackage{graphicx}	
\usepackage[font=small]{caption}
\usepackage{subcaption}
\usepackage{longtable}
\usepackage{rotating}
\usepackage{hyperref}
\usepackage{tikz}
\usepackage{color}
\usepackage{soul}
\usepackage{array}
\usepackage{multirow}
\usepackage{geometry}
\usepackage{natbib}
\usepackage[nolist]{acronym}
\usepackage{arydshln}
\usepackage{float}

\usepackage[percent]{overpic} 

\newcommand{\Ha}{H$\alpha$}

\newcommand{\HI}{H{\sc i}}
\def\HII{\hbox{H{\sc ii}}}
\newcommand{\SII}{[S{\sc ii}]}

\newcommand{\OIII}{[O{\sc iii}]}
\newcommand{\D}{$^\circ$}

\def\arcmin{\hbox{$^\prime$}}
\def\arcsec{\hbox{$^{\prime\prime}$}}
\def\kms{km\,s$^{-1}$}

\def \xmm {\emph{XMM-Newton}}
\newcommand{\mjybm}{\,mJy\,beam$^{-1}$}

\begin{acronym}
\acro{2MASS}{The Two Micron All-Sky Survey}
\acro{30Dor}[30 Dor]{30~Doradus}
\acro{AGN}{active galactic nucleus}
\acrodefplural{AGN}{Active Galactic Nuclei}
\acro{ATCA}{Australia Telescope Compact Array}
\acro{ATNF}{Australia Telescope National Facility}
\acro{ATOA}{Australia Telescope Online Archive}
\acro{AT20G}{Australia Telescope 20 GHz Survey}
\acro{ASKAP}{Australian Square Kilometre Array Pathfinder}
\acro{BL}[BL-Lac]{BL Lacertae Objects}
\acro{CASS}{CSIRO Astronomy and Space Science}
\acro{CABB}{Compact Array Broadband Back-end}
\acro{CHII}[{\sc CHii}]{Compact \textsc{Hii}}
\acro{CSIRO}{Australian Commonwealth Scientific and Industrial Research Organisation}
\acro{CSS}{Compact Steep Spectrum}
\acro{CME}{Coronal Mass Ejection}
\acro{Dec}{Declination}
\acro{DD}{double-degenerate}
\acro{DS9}[\textsc{DS9}]{\textsc{SAOImage DS9}}
\acro{DSS}{Digital Sky Survey}
\acro{EMU}{Evolutionary Map of the Universe}
\acro{ev}[eV]{electronvolt\acroextra{: 1 eV $\approx 1.6 \times 10^{-19}$ J}}
\acro{FITS}[\textsc{Fits}]{Flexible Image Transport System}
\acro{FRB}{Fast Radio Bursts}
\acro{FRII}{Fanaroff-Riley II}
\acrodefplural{FRII}{Fanaroff-Riley II's}
\acro{FRI}{Fanaroff-Riley I}
\acrodefplural{FRI}{Fanaroff-Riley I's}
\acro{FSRQ}{Flat Spectrum Radio Quasars}
\acro{FWHM}{full width at half-maximum}
\acro{GLEAM}{GaLactic Extragalactic All-sky MWA}
\acro{GMRT}{Giant Meterwave Radio Telescope}
\acro{GRG}{Giant Radio Galaxy}
\acro{GPS}{Gigahertz Peak Spectrum}
\acro{HEMT}{High Electron Mobility Transistor}
\acro{HFP}[HFP]{High Frequency Peaker}
\acro{HzRGs}{High Redshift Radio Galaxies}
\acrodefplural{HVS}{hypervelocity stars}
\acro{HVS}{hypervelocity star}
\acro{pHFP}[pHFP]{Potential High Frequency Peaker}
\acro{HST}{\textit{Hubble Space Telescope}}
\acro{IAU}{International Astronomical Union}
\acro{IFRSs}{Infrared Faint Radio Sources}
\acro{ISM}{interstellar medium}
\acro{IR}{Infrared}
\acro{JY}[Jy]{Jansky\acroextra{, 1 Jy = $10^{-26} \times \mathrm{W~ m}^{-2}~\mathrm{Hz}^{-1}$}} 
\acro{LLS}{Largest Linear Size}
\acro{LFAA}{Low-Frequency Aperture Array}
\acro{LMC}{Large Magellanic Cloud}
\acro{LG}{Local Group}
\acro{LSO}[LSOs]{Large Scale Objects}
\acro{MACHO}{Massive Astrophysical Compact Halo Objects}
\acro{MC}[MCs]{Magellanic Clouds}
\acro{mc}[MC]{Magellanic Cloud}    
\acro{MCELS}{Magellanic Cloud Emission Line Survey}
\acro{MW}{Milky Way}
\acro{MIRIAD}[\textsc{Miriad}]{Multichannel Image Reconstruction, Image Analysis and Display}
\acro{MIT}{Massachusetts Institute of Technology}
\acro{MOST}{Molonglo Observatory Synthesis Telescope}
\acro{MQS}{Magellanic Quasars Survey}
\acro{MRC}{Molonglo Reference Catalogue of Radio Sources}
\acro{MWA}{Murchison Widefield Array}
\acro{NAT}{Narrow-Angle Tail}
\acro{NRAO}{National Radio Astronomy Observatory}
\acro{NVSS}{NRAO VLA Sky Survey}
\acrodefplural{ORC}{Odd Radio Circles}
\acro{ORC}{Odd Radio Circle}
\acro{OPAL}{Online Proposal Applications \& Links}
\acro{OVV}{Optically Violent Variable Quasars}            
\acro{PAF}{Phased Array Feed}
\acro{pc}{parsec\acroextra{: 1 pc $\simeq 3.09 \times 10^{16}$ m}}
\acro{PMN}{Parkes-MIT-NRAO}
\acro{PNe}{Planetary Nebulae}
\acro{PN}{Planetary Nebula}
\acro{PWN}{Pulsar Wind Nebula}
\acro{QSO}{Quasi-Stellar Object}
\acro{RA}{Right Ascension}
\acro{RFI}{Radio-Frequency Interference}
\acro{RMS}[rms]{Root Mean Squared}
\acro{SDSS}{Sloan Digital Sky Survey}
\acro{SED}{spectral energy distribution}
\acro{SI}[$\alpha$]{Spectral Index\acroextra{, $S \propto \nu^\alpha$}}
\acro{SKA}{Square Kilometre Array}
\acro{SMASH}{Survey of the MAgellanic Stellar History}
\acro{SMB}{Super Massive Blackholes}
\acro{SMC}{Small Magellanic Cloud}
\acro{SN}{Supernova}
\acro{SNe}{Supernovae}
\acro{SNR}{Supernova Remnant}
\acrodefplural{SNR}{Supernova Remnants}
\acro{SD}{single-degenerate}
\acro{SUMSS}{Sydney University Molonglo Sky Survey}
\acro{SMBH}{Super Massive Black Hole}
\acro{TOPCAT}[\textsc{Topcat}]{Tool for OPerations on Catalogues And Tables}
\acro{USS}{Ultra Steep Spectrum}
\acro{WBAC}{Wide-Band Analogue Correlator}
\acro{WIFES}[WiFeS]{Wide-Field Spectrograph}
\acro{WISE}{Wide-Field Infrared Survey Explorer}
\acro{VLA}{Very Large Array} 
\acro{VLBI}{Very Long Baseline Interferometry} 
\acro{VLSR}[\textbf{$v_{lsr}$}]{Velocity in the Line of Sight}
\acro{WD}{white dwarf}
\end{acronym}

\title[New ASKAP Radio SNRs and candidates in the LMC]{New ASKAP Radio Supernova Remnants and Candidates in the Large Magellanic Cloud }

\author[Bozzetto, L. M., Filipovic, M. D.]
{Luke M. Bozzetto,$^{1}$ 
Miroslav D.~Filipovi\'c$^{1}$,\thanks{E-mail: m.filipovic@westernsydney.edu.au}
H. Sano,$^{2}$
R. Z. E. Alsaberi,$^{1}$ 
L. A. Barnes$^{1}$
\newauthor
I. S. Boji\v ci\'c,$^{1}$
R. Brose,$^{3}$
L. Chomiuk,$^{4}$
E. J. Crawford,$^{1}$
S. Dai,$^{1}$
M.~Ghavam,$^{1}$
\newauthor
F. Haberl,$^{5}$
T. Hill,$^{1}$
A. M. Hopkins,$^{6,1}$
A. Ingallinera,$^{7}$ 
T. Jarrett,$^{8,1}$  
P.~J.~Kavanagh,$^{9}$
\newauthor
B. S. Koribalski,$^{10,1}$
R. Kothes,$^{11}$
D. Leahy,$^{12}$
E. Lenc,$^{10}$
I. Leonidaki,$^{13}$
P. Maggi,$^{14}$
\newauthor
C. Maitra,$^{5}$
C.~Matthew,$^{1}$
J. L.~Payne,$^{1}$
C. M. Pennock,$^{15}$
S. Points,$^{16}$ 
W. Reid,$^{1,17,18}$  %
\newauthor
S. Riggi,$^{7}$  
G. Rowell,$^{19}$
M. Sasaki,$^{20}$ 
S. Safi-Harb,$^{21}$ 
J. Th. van Loon,$^{10}$
N. F. H. Tothill,$^{1}$
\newauthor
D. Uro\v sevi\' c,$^{22,23}$
F. Zangrandi,$^{20}$
\\
\\
Affiliations are listed at the end of the paper
}

\date{Accepted XXX. Received YYY; in original form ZZZ}

\pubyear{2022}

\begin{document}
\label{firstpage}
\pagerange{\pageref{firstpage}--\pageref{lastpage}}
\maketitle

\begin{abstract}
We present a new \ac{ASKAP} sample of 14 radio \ac{SNR} candidates in the \ac{LMC}. This new sample is a significant increase to the known number of older, larger and low surface brightness \ac{LMC} \acp{SNR}. 
We employ a multi-frequency search for each object and found possible traces of optical and occasionally X-ray emission in several of these 14 \ac{SNR} candidates. 
One of these 14 \ac{SNR} candidates (MCSNR~J0522--6543) has multi-frequency properties that strongly indicate a bona fide \ac{SNR}.
We also investigate a sample of 20 previously suggested \ac{LMC} \ac{SNR} candidates and confirm the \ac{SNR} nature of MCSNR~J0506--6815.
We detect lower surface brightness \ac{SNR} candidates which were likely formed by a combination of shock waves and strong stellar winds from massive progenitors (and possibly surrounding OB stars).
Some of our new \ac{SNR} candidates are also found in a lower density environments in which SNe type~Ia explode inside a previously excavated \ac{ISM}. 
\end{abstract}




\section{Introduction}
 \label{sec:intro}

\ac{SNe} are powerful explosions that mark the end for certain types of stars. The study of \ac{SNe} and their remnants is essential to understanding the physical and chemical evolution of the \ac{ISM}. \ac{SNe} are typically categorised as arising either from a core-collapse event occurring in more massive, short-lived stars ($\geq$8\,M$_\odot$), or a thermonuclear type~Ia event occurring in carbon-oxygen white dwarfs. Both types enrich the \ac{ISM} with heavy elements, create a shock wave that heats the swept-up \ac{ISM}, compress magnetic fields, and accelerate particles such as cosmic rays \citep[e.g.][]{2021pma..book.....F}. 

In the search for the imprints of these \ac{SNe} explosions, a few key characteristics are used to classify an object as a \ac{SNR}: a non-thermal radio spectral index of $\alpha<-0.4$ (defined as $S_{\nu}$~$\propto$~$\nu^\alpha$, where $S_{\nu}$ is flux density, $\nu$ is frequency and $\alpha$ is the spectral index), diffuse X-ray emission and an elevated \SII/\Ha\ ratio $\geq0.4$ which is produced by the high-velocity shocks \citep[e.g.][]{2021map..book....2H,2021map..book....6M,2021MNRAS.500.2336Y}. Not all \acp{SNR} exhibit all three of these characteristics. Typically, two of these characteristics are enough to confirm a \ac{SNR} classification, while one of them marks the source as an \ac{SNR} candidate \citep{2012SSRv..166..231R,2016A&A...585A.162M,2017ApJS..230....2B,2019A&A...631A.127M}. 
Also, a variety of \ac{SNR} types (such as \ac{PWN}) produces other recognisable signatures such as the polarised radio emission and flatter radio spectral index.

The \ac{LMC}, an irregular dwarf galaxy, has been the target of extensive studies into \acp{SNR} since they were first observed by \citet{mathewson_healey_1964}. Details of the \ac{LMC} \ac{SNR} studies are covered in \citet{2017ApJS..230....2B} \& \citet{2021MNRAS.500.2336Y}. The \ac{LMC} is of particular interest in the hunt for \acp{SNR} because of its low foreground absorption and relatively close proximity of 50~kpc \citep{2019Natur.567..200P}. While the proximity of the \ac{LMC} allows us to study our nearest Galactic neighbour in more detail than those further away, for all intents and purposes, all objects within the \ac{LMC} are assumed to be at the same approximate 50~kpc. 

The \ac{LMC} also hosts a mini-starburst at sub-solar metallicity. This may result in a dis-proportionally high number of very massive stars  \citep[$>$100~M$_{\odot}$;][]{2018Sci...359...69S}
and pair-instability \ac{SNe} which ultimately produce \acp{SNR}. The mean energy of the  \ac{LMC} \acp{SNR} is $\sim5\times10^{50}$~erg, very similar to Galactic \acp{SNR} \citep[$\sim4\times10^{50}$~erg;][]{2017ApJ...837...36L,2020ApJS..248...16L}.

\begin{figure*}
 \begin{center}
    \includegraphics[width=\textwidth,trim = 100 0 105 0,clip]{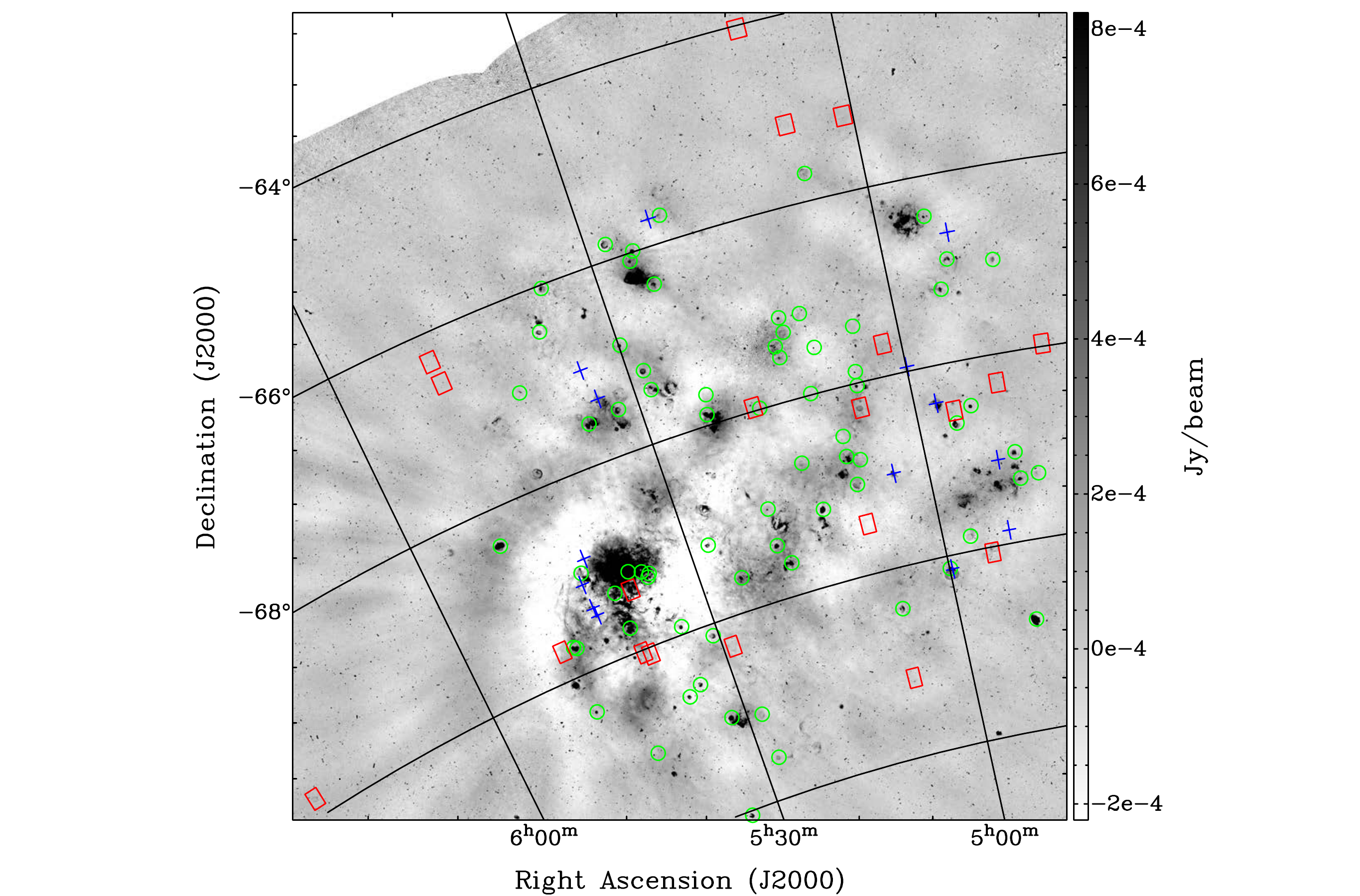}
    \caption{14 new \ac{SNR} candidate positions in the \ac{LMC} are marked in blue crosses while 71 green circles represent known \acp{SNR} and 20 red rectangles are previously established \ac{SNR} candidates. The background image is from the \ac{ASKAP} survey of the \ac{LMC} and the scale bar on the right hand side shows the surface brightness intensity scale in Jy~beam$^{-1}$.} 
 \label{fig:1a}
 \end{center}
\end{figure*}

The new generation of radio telescopes such as \ac{ASKAP} and MeerKAT, with their higher sensitivity instrumentation, give us an unprecedented look into some of the older, larger, faded remnants -- especially those that are embedded or obscured by nearby stronger sources \citep{2021PASA...38....3N,2021MNRAS.505L..11K,2022MNRAS.512..265F,2022MNRAS.509.5209J} and those expending into cavities blown by the progenitor. Certainly, small size \ac{LMC} \acp{SNR} ($<15$~arcsec or 3.5~pc at the \ac{LMC} distance) are presumably young (and therefore bright) and they are most likely discovered in the previous surveys \citep{2022PASP..134f1001U}.

In this paper, we make use of new radio continuum data taken by the \ac{ASKAP} telescope \citep{2021MNRAS.506.3540P,2021MNRAS.507.2885F} to add 14 new candidates to the \ac{LMC} \ac{SNR} sample, creating the most complete \ac{SNR} sample of any galaxy (see Fig.~\ref{fig:1a}). Moving towards such a comprehensive population is essential for statistical analysis and the evolutionary understanding of the remains of exploded stars. A complete sample of \acp{SNR} provides the ability to study their global properties, in addition to carrying out detailed analyses of their sub-classes (e.g., sorted by X-ray, optical and radio morphology or by progenitor \ac{SN} type).

\section{Observations and Data Processing}

This paper draws together various existing surveys of the \ac{LMC} at multiple wavelengths to identify new candidate \acp{SNR}, with a primary focus on the sources identified in the new \ac{ASKAP} observations of \citet[][see Fig.~\ref{fig:1a}]{2021MNRAS.506.3540P}.

\subsection{ASKAP data}
These \ac{ASKAP} radio data comprise a twelve-pointing mosaic taken of the \ac{LMC} at a centre frequency of 888~MHz (bandwidth of 256~MHz, spatial resolution of 13.9$\times$12.1~arcsec$^2$ and position angle $-84\degr$). Details of the observations, data reduction and analysis of these data have been presented in \citet{2021MNRAS.506.3540P}. The image  shown in Fig.~\ref{fig:1a} is affected by various artefacts and missing short spacing data. However, its robust maximum angular scale of 49~arcmin combined with improved sensitivity \citep[][]{2022MNRAS.512..265F} ensures a reasonable environment to search for new \ac{LMC} \acp{SNR}. At the time of these observations (April~2019) \ac{ASKAP} didn't have polarisation capability, and therefore, the \ac{ASKAP} image used here is total intensity (Stokes~I).

\subsection{ATCA data}
We observed MCSNR~J0522--6543\footnote{To distinguish between \ac{SNR} candidates and confirmed \acp{SNR}, \citet{2016A&A...585A.162M} established nomenclature where bona fide \ac{MC} \acp{SNR} are named with the prefix `MCSNR~J' and candidates with only `J'.} with the \ac{ATCA} on 30$^\mathrm{th}$ November 2019 and 23$^\mathrm{rd}$ February 2020 (project codes C3275 and C3292; see Appendix~\ref{sec:0522notes}). The observations were carried out in frequency switching mode (between 2100~MHz and 5500/9000~MHz) with 1-hour of integration over a 12-hour period using arrays 1.5C and EW367. The \ac{CABB} with its 2048\,MHz bandwidth was used centred at $\nu$=5500 and 9000\,MHz, totaling 117.62~minutes of integration. At the same time, we also used the $\nu$=2100\,MHz band with a total of 112.2~minutes of integration. The primary (flux) calibrator was PKS\,B1934--638 ($S_{2100}$=11.651, $S_{5500}$=5.010 and $S_{9000}$=2.704~Jy) and secondary calibrator (phase) was PKS\,B0530--727 ($S_{2100}$=0.680, $S_{5500}$=0.585 and $S_{9000}$=0.695~Jy). For imaging, we used \textsc{WSClean} \citep{offringa-wsclean-2014} and flagged out data from the 6$^{\rm th}$ antenna for all observations to focus on the extended emission of this object at the expense of reducing the resolution. We achieved a resolution of 20.95$\times$16.60~arcsec$^2$ at 2100~MHz, 15.85$\times$12.79~arcsec$^2$ at 5500~MHz and 8.81$\times$7.05~arcsec$^2$ at 9000~MHz (all at P.A.=0 degrees) with a corresponding rms noise of 0.1, 0.05 and 0.025~\mjybm, respectively. 

\subsection{Optical, X-ray and IR data}
The optical data used in this study consists of \Ha, \SII, \OIII\ images from the \ac{MCELS} survey; more details can be found in \citet{2019ApJ...887...66P}. 

To gain insight into the type of stellar environment hosting the suspected progenitors of these objects, we use data sourced from the Magellanic Clouds Photometric Survey \citep[MCPS;][]{2004AJ....128.1606Z}. To this end, we construct colour-magnitude diagrams and identify blue stars more massive than $\sim$8\,M$_{\sun}$ within a 100~pc (6.9~arcmin at the distance of the \ac{LMC}; $\sim$10$^{7}$~yr at 10~\kms) radius of each of the objects presented in this study. The size of \acp{SNR} can well exceed 150~pc in diameter and therefore our chosen search area will give us a good view of the environment and possible origin of these objects as demonstrated in \citet[][]{2017ApJS..230....2B}. This allows us to see the prevalence of early-type stars close to the candidate remnants.

We also consult various X-ray surveys including {\it ROSAT} \citep{1999A&AS..139..277H} and \xmm\ \citep{2019svmc.confE..63H}. Finally, we use the Spitzer infrared survey of the \ac{LMC}, ``Surveying the Agents of a Galaxy's Evolution'' (SAGE) \citep{2006AJ....132.2268M}.

\subsection{\HI\ data}

For atomic hydrogen, we used the \ac{ATCA} \& Parkes \HI\ survey data from \citet{2003ApJS..148..473K} with its angular resolution of 60~$\times$~60~arcsec$^2$. 
We made the velocity channel maps and position-velocity ($p-v$) diagrams to constrain the velocity ranges of \HI\ clouds that physically interact with the \acp{SNR}. The velocity channel maps provide a spatial correspondence between the \ac{SNR} shell boundary and \HI\ clouds as tested by previous studies \citep[e.g.][]{2005ApJ...631..947M}. When the \HI\ clouds are associated with the \ac{SNR}, they are expected to be located along with the shell. Likewise, \HI\ cavities in $p-v$ diagrams are also expected when supernova shocks and/or stellar winds from the progenitor evacuated (accelerated) the \HI\ gas.

In order to define the shock-interacting velocity range of \HI\ clouds, we first prepared velocity channel maps of \HI\ towards each \ac{SNR} candidate.
Because the shock-interacting clouds will be limb-brightened in the synchrotron radio continuum through the shock-cloud interactions with the magnetic field amplification, the velocity channel distribution of \HI\ is useful to identify the spatial correlation between the shocked \HI\ clouds and \ac{SNR} candidate shells.

Here, we prepared a velocity channel distribution of \HI\ as a velocity step of 6.6~\kms\ (see Appendix~\ref{sec:snrs}). The map was centred at each \ac{SNR} candidate with a size of $\sim$50~$\times$~50~arcmin$^2$. 
Finally, we searched for expanding shell-like structures in the $p-v$ diagram by changing the integration range of R.A. When the expanding gas motion is identified, we selected the \HI\ velocity range which traces the inner edge of the cavity in the $p-v$ diagram.

\section{Results \& Discussion}
\label{sec:results}

The first extragalactic \ac{SNR} was discovered over half a century ago as the \ac{LMC} \ac{SNR} N\,49 \citep{1963Natur.199..681M}. Ever since then, studies of our \ac{MW} immediate neighbouring galaxy the \ac{LMC} \ac{SNR} sample is recognised as of essential importance because it provides the best opportunity to reach a complete sample of these objects. With every new generation of astronomical instruments, we improved our knowledge about these objects as they come in large varieties. One of the most important aspects of these previous studies is to detect new \acp{SNR} which in return provides a better view of their evolutionary processes in various environments and with different progenitors.

As introduced in Section~\ref{sec:intro}, we divide the \ac{LMC} \ac{SNR} population into two groups: bona fide \acp{SNR} and \ac{SNR} candidates. Previous studies of the \ac{LMC} \acp{SNR} revealed {71 confirmed objects \citep[58+1+1+1+3+7 as in][]{2017ApJS..230....2B,2016A&A...585A.162M,2019MNRAS.490.5494M,2021MNRAS.504..326M,2021MNRAS.500.2336Y,2022MNRAS.tmp..786K} and 20 candidates \citep[4+15+1 as in][]{2017ApJS..230....2B,2021MNRAS.500.2336Y,2022MNRAS.512..265F}}. These studies are mainly based on the previous generation of \ac{ATCA} radio, \xmm\ and {\it Chandra} X-ray and \ac{MCELS} optical surveys. One of the main conclusions from \citet{2017ApJS..230....2B} study of the \ac{LMC} \ac{SNR} sample is that we were missing detection of large size but low surface brightness \acp{SNR}. It was also acknowledged that some \acp{SNR} could escape detection as they could be embedded into the large scale \HII\ regions such as 30~Doradus \citep{2015A&A...573A..73K}. Along that expectation, we endeavour in a search for new \ac{LMC} \acp{SNR} using the latest generation of \ac{ASKAP} radio survey \citep{2021MNRAS.506.3540P}.

\subsection{Previous LMC SNR candidates}

We first investigate the present \ac{LMC} \ac{SNR} candidate sample of 20 objects. \citet{2021MNRAS.500.2336Y} found that none of their 15 optically selected candidates can be detected in our \ac{ASKAP} survey. In addition to these, \citet{2022MNRAS.512..265F} suggested that ASKAP~J0624--6948 is an intergalactic \ac{SNR} positioned in the outskirts of the \ac{LMC}, but the true nature of this object remains mysterious. 

\begin{figure*}
  \begin{center}
    \includegraphics[width=\textwidth,trim = 0 0 0 0,clip]{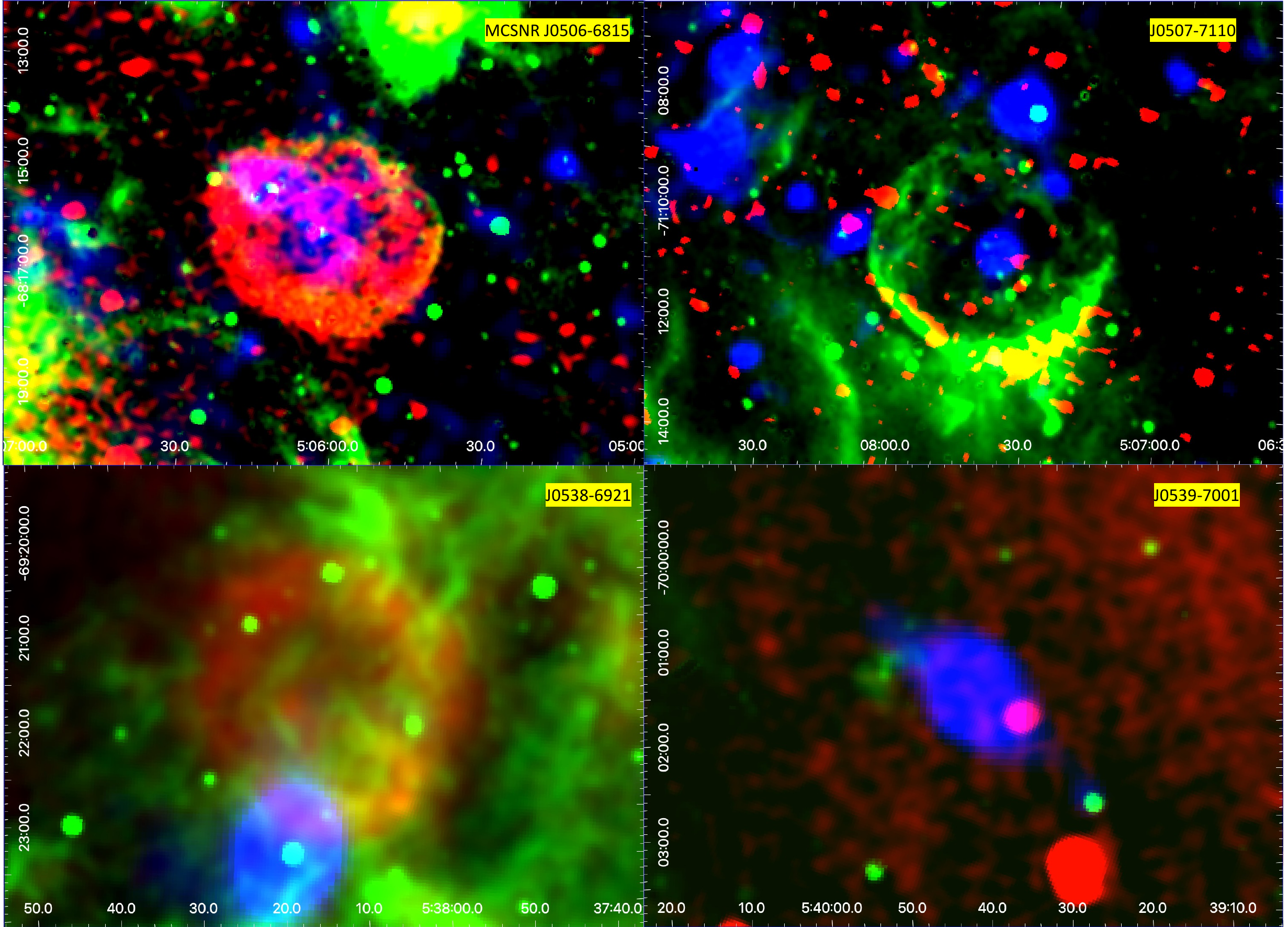}
\caption{Previously selected \ac{LMC} \ac{SNR} candidates MCSNR~J0506--6815 (top left), J0507--7110 (top right), J0538--6921 (bottom left) and J0539--7001 (bottom right). All RGB images  made from the \ac{ASKAP} 888~MHz radio data are in red (at 13.9$\times$12.1~arcsec$^2$ resolution), \Ha\ in green and \xmm\ at 0.2--1.0~keV in blue. The previously selected \ac{SNR} candidate, J0506--6815, is confirmed as a bona fide \ac{LMC} \ac{SNR}.}
    \label{fig:4cand}
  \end{center}
\end{figure*}

This leaves four remaining \ac{LMC} \ac{SNR} candidates that are initially investigated in \citet{2017ApJS..230....2B}: J0506--6815 ([HP99]\,635), J0507--7110 (DEM\,L81), J0538--6921 (MC\,73) and J0539--7001 ([HP99]\,1063)). We search for their radio continuum \ac{SNR} signature in our \ac{ASKAP} survey (see Fig.~\ref{fig:4cand}) as well as in other wavelengths. 

\begin{figure}
  \begin{center}
    \includegraphics[width=0.65\columnwidth,trim = 50 100 0 0,angle=-90,clip]{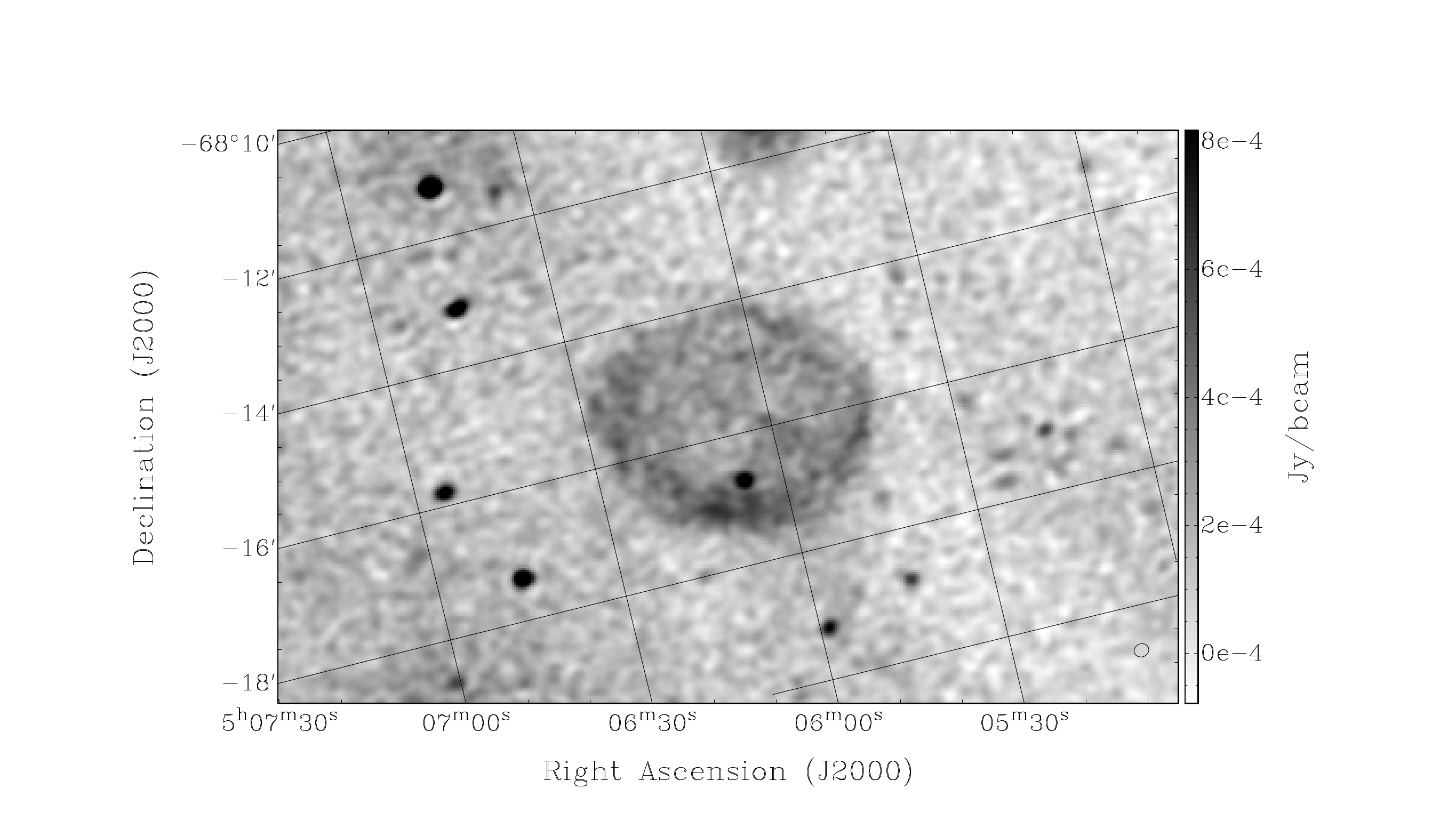}
\caption{\ac{ASKAP} image of MCSNR~J0506--6815 with a beam size of 13.9$\times$12.1~arcsec$^2$ (bottom right corner). We note unrelated point source towards the southern edges of the \ac{SNR}.} 
    \label{fig:0506askap}
  \end{center}
\end{figure}

We find that J0506--6815 ([HP99]\,635) is a circular source in our \ac{ASKAP} radio image (see Figs.~\ref{fig:4cand} (top left panel) and \ref{fig:0506askap}) with an estimated flux densities of $S_{\rm 888\,MHz}$=78$\pm$2~mJy and $S_{\rm 1377\,MHz}$=64$\pm$4~mJy (from \citet{2021MNRAS.507.2885F}) produce a spectral index $\alpha=-0.45\pm0.24$. Together with a prominent central soft X-ray emission in the \xmm\ survey and \SII/\Ha$>$0.9, we now safely confirm MCSNR~J0506--6815 as a bona fide \ac{SNR}. Interestingly, the X-ray emission occupies central part of the \ac{SNR} while radio and optical emission dominates at the edges. This anti-correlation indicates that MCSNR~J0506--6815 could be very similar to the known iron-rich cores \acp{SNR} such as MCSNR~J0506--7025 and MCSNR~J0527--7104 \citep{2016A&A...586A...4K,2022MNRAS.tmp..786K}.

\ac{SNR} candidate J0507--7110 (DEM\,L81; Fig.~\ref{fig:4cand} top right panel) shows some typical \ac{SNR} morphological characteristics, but in our \ac{ASKAP} \ac{LMC} radio image this object is confused by the nearby emission and other sources. Therefore, we are keeping its status as a candidate \ac{SNR}. 

J0538--6921 (MC\,73; Fig.~\ref{fig:4cand} bottom left panel) is certainly a prototype radio \ac{SNR} candidate based on its radio properties ($S_{\rm 888\,MHz}$=294$\pm$3~mJy; $S_{\rm 1377\,MHz}$=248$\pm$4~mJy; $S_{\rm 4800\,MHz}$=120$\pm$10~mJy; $\alpha=-0.56\pm0.02$) but lack of X-ray or optical confirmation prevent us from a final classification. 

We clearly see a point radio source close to the centre of X-ray \ac{LMC} \ac{SNR} candidate J0539--7001 ([HP99]\,1063; Fig.~\ref{fig:4cand} bottom right panel), but we cannot confirm its \ac{SNR} nature as the positional displacement between radio and X-ray position of $\sim$25~arcsec is not negligible. Also, no other nearby and obvious fine-scale structure could be linked with the object's possible \ac{SNR} nature. However, we acknowledge that a compact type of \ac{SNR} candidate such as J051327--6911 \citep{2007MNRAS.378.1237B} is still a possibility.

\subsection{New ASKAP LMC SNR candidates}
We searched our new \ac{ASKAP} images of the \ac{LMC} for circular shaped objects (above 5$\sigma$ of local noise) with an enhanced ratio of radio continuum to \Ha\ emission. This method, introduced by \citet{1991MNRAS.249..722Y}, successfully identifies objects in which non-thermal emission is dominant, i.e. \acp{SNR}. Using this enhanced ratio, combined with the typically circular \ac{SNR} morphology, we find 14 regions in which the presence of an \ac{SNR} is plausible. We present a catalogue of radio images of these newly discovered \ac{SNR} candidates in Fig.~\ref{fig:X} and details in Appendix~\ref{sec:snrs}.

\begin{figure*}
  \begin{center}
    \includegraphics[width=\textwidth,trim = 0 100 0 100,clip]{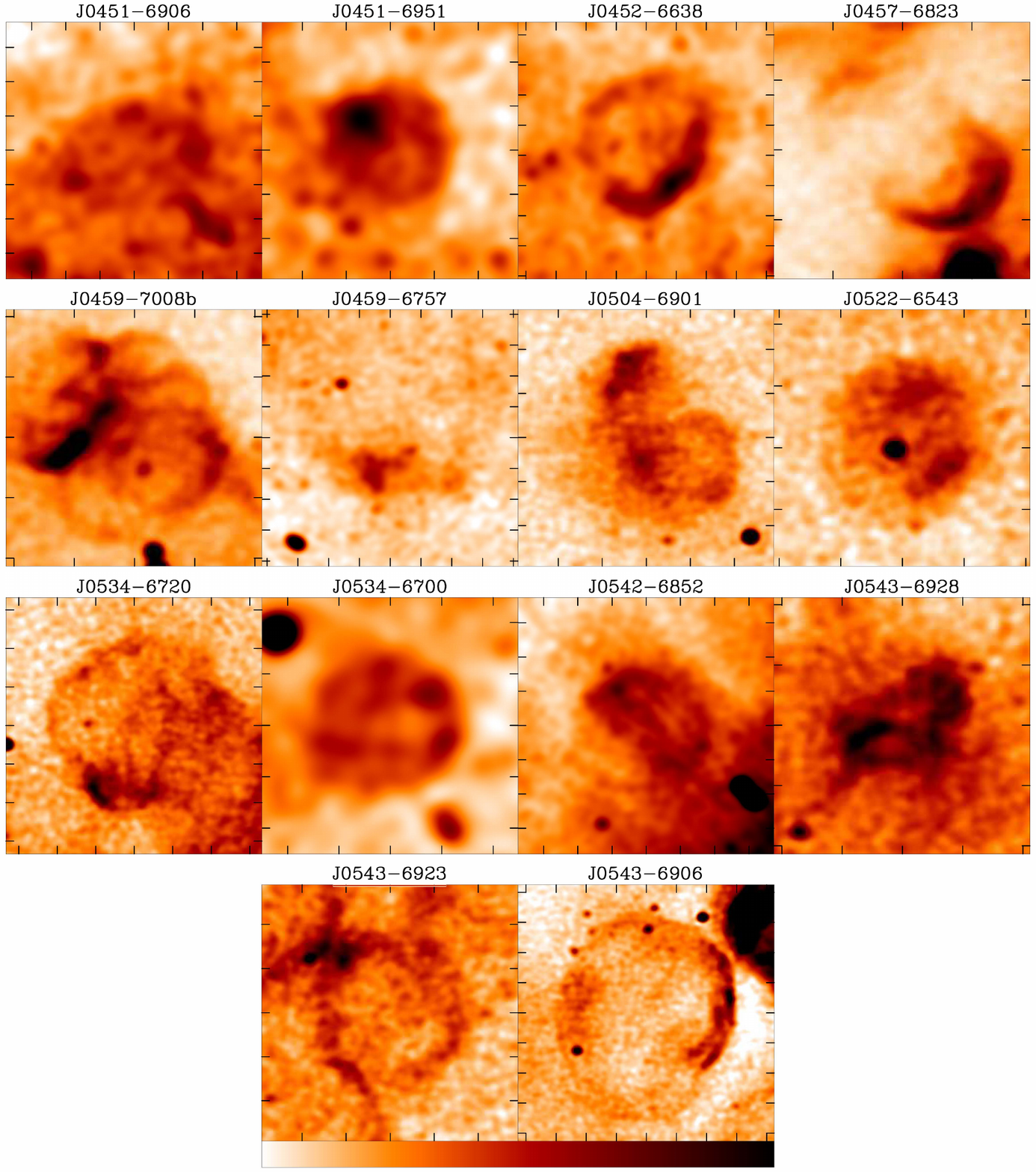}
    \caption{\ac{ASKAP} radio images of all 14 newly discovered \ac{LMC} \acp{SNR} candidates centred at the locations indicated in Table~\ref{tab:snrtable}. Tick intervals of 1~arcmin  correspond to 14.5~pc at the distance of the \ac{LMC}. The images for \ac{SNR} candidates J0459--6757 and J0542--6852 have been convolved to 20~arcsec, those of \ac{SNR} candidates J0451--6906, J0451--6951, J0452--6638 and J0534--6700 to 30~arcsec in order to enhance the signal-to-noise ratio of low surface brightness sources. The colour bar for all images is shown at the bottom of the panel. The intensity ranges in mJy/beams for the 14 panels are from top left to bottom right are: ($-$0.2 : 1.6) ($-$0.8 : 0.5) ($-$0.8 : 1.1) ($-$0.2 : 3.3) ($-$0.2 : 2.5) ($-$0.2 : 1.7) ($-$0.2 : 1.5) ($-$0.2 : 1.2) ($-$0.2 : 1.2) ($-$1.1 : 0.6) ($-$1.4 : 1.2) ($-$0.4 : 1.0) ($-$0.4 : 1.0) ($-$0.3 : 1.5). We note that J0522--6543 is confirmed \ac{SNR} in this study.}
    \label{fig:X}
  \end{center}
\end{figure*}

We used the method described in \citet[][Section 2.4]{2019PASA...36...48H} and \citet[][]{2022MNRAS.512..265F} to measure the position, extent and flux densities of all 14 selected \ac{LMC} \ac{SNR} candidates. In the same manner, we measure the three \ac{ATCA} observations of MCSNR~J0522--6543 (at 2100, 5500 and 9000~MHz). In short, we carefully selected proposed \ac{SNR} regions that exclude all obvious point sources and then measured the total radio flux density, accounting for local background. We note the selected \ac{SNR} candidates have low surface brightness and/or are sometimes embedded in complex environments (c.f. J0457--6823, J0459--7008b\footnote{Because of the close proximity (and similar position) to \ac{SNR} N\,186D, we use `b' (as J0459--7008b) to distinguish between two possibly separate objects. } and J0459--6757). This significantly influences the accuracy or may prevent any meaningful measurements. We estimate our flux density measurements have an error of $<$20~per~cent. Also, \citet[][see their Figure~7]{2022MNRAS.512..265F} noted that a lack of short spacing data in observations\footnote{We also used the same 888~MHz image and the same \ac{ATCA} array observations with the similar integration time.} may result in under-estimates of extended flux density measurements. \citet{2022MNRAS.512..265F} suggest that spectral index estimates could be as much as $\sim$15--20\,\% (or $\alpha\sim$0.1) flatter.

In Table~\ref{tab:snrtable} we list the source name in Column~1, its central position (RA and DEC) in Columns~2 and 3, source angular extent as major and minor axis/diameter in Column~4 (in arcsec), arithmetic average of major and minor axes/diameter converted to parsecs for a distance of 50~kpc in Column~5, and position angle (PA) in Column~6, measured from north to east. 
Our flux density estimates at 888~MHz are listed in Column~7. As none of the here selected \ac{LMC} \ac{SNR} candidates (except for MCSNR J0522--6543) can be seen at other radio frequencies, we can not estimate their spectral index. To compare the surface brightness of our sample with established \ac{LMC} \acp{SNR}, we assumed a typical \ac{SNR} spectral index of $\alpha=-0.5$ \citep{2012SSRv..166..231R, 2014SerAJ.189...15G,2017ApJS..230....2B,2019A&A...631A.127M,book2}. However, we acknowledge that some of our \ac{SNR} candidates may contain an old \ac{PWN} which would have a somewhat flatter spectral index. Using this assumed spectral index allowed us to estimate the flux density and surface brightness of these sources at 1~GHz (Column~8; $\Sigma_{\rm 1\,GHz}$). 
In Column~9 we provide the number of massive OB stars found within a $\sim$100~pc radius of each of the objects' central position. Lastly, in Columns~10 and 11, Y/N apply when optical and X-ray data are available and/or the source is detected.

\begin{table*}
	\centering
	\caption{The main properties of the 14 new \ac{LMC} \ac{SNR} candidates found in this study. The positional accuracy measurements are better then 2~arcsec while flux density errors are $<$20~per~cent. $\dag$ indicates that J0522--6543 is confirmed \ac{SNR} in this study.
	}
	\label{tab:snrtable}
	\begin{tabular}{lccccrccccc} 
		\hline
Name    & RA (J2000) & DEC (J2000)             & $\theta_{\rm maj} \times \theta_{\rm min}$  & $D_{\rm av}$  & PA    & $S_{\rm 888\,MHz}$ & $\Sigma_{\rm 1\,GHz}$ & \# OB Star    & Optical   & X-ray \\
        & (h m s)    &  (\D\ \arcmin\ \arcsec) & (arcsec)                            & (pc)          & (\D)  & (mJy)              & (W~m$^{-2}$~Hz$^{-1}$sr$^{-1}\times10^{-22}$) & Candidates  & ID & ID \\
		\hline
	\noalign{\smallskip}
J0451--6906     & 04 51 38.9& --69 06 26 & 299 $\times$ 194 & 58 & 90   & 27 & 2.41 & 19 & Y & -- \\
J0451--6951     & 04 51 50.8& --69 51 30 & 170 $\times$ 168 & 41 & 0    & 1 & 0.21 & 10 & N & N \\
J0452--6638     & 04 52 42.4& --66 38 35 & 277 $\times$ 196 & 56 & 90   & 5 & 0.04 & 29 & Y & -- \\
J0457--6823     & 04 57 30.9& --68 23 35 & 196 $\times$ 115 & 36 & 60   & ---   & ---  & 33 & Y & ? \\
J0459--7008b    & 04 59 38.7& --70 08 37 & 203 $\times$ 178 & 46 & 150  & ---   & ---  & 44 & Y & -- \\

J0459--6757     & 04 59 54.9& --67 57 04 & 131 $\times$ 114 & 30 & 0    & ---   & ---  & 18 & Y & -- \\
J0504--6901     & 05 04 04.8& --69 01 12 & 259 $\times$ 246 & 61 & 22   & 93 & 7.49 & 72 & Y & ? \\
J0522--6543$\dag$     & 05 22 53.5& --65 43 09 & 171 $\times$ 159 & 40 & 0    & 35 & 6.54 & 15 & Y & -- \\ 
J0534--6720     & 05 34 05.5& --67 20 48 & 293 $\times$ 288 & 70 & 0    & 102 & 6.15 & 45 & N & -- \\
J0534--6700     & 05 34 48.7& --67 00 01 & 188 $\times$ 179 & 45 & 0    & 1 & 0.21 & 325& N & --\\

J0542--6852     & 05 42 05.9& --68 52 14 & 251 $\times$ 218 & 57 & 75   & 74 & 6.92 & 40 & Y & Y \\
J0543--6928     & 05 43 06.3& --69 28 42 & 157 $\times$ 104 & 31 & 125  & 45 & 14.11& 34 & N & N \\
J0543--6923     & 05 43 16.5& --69 23 27 & 228 $\times$ 213 & 54 & 0    & 28 & 2.92 & 31 & ? & N \\
J0543--6906     & 05 43 25.2& --69 07 19 & 443 $\times$ 354 & 96 & 0    & 83 & 2.42 & 50 & N & ? \\
	\noalign{\smallskip}
		\hline
	\end{tabular}
\end{table*} 

These 14 new \ac{SNR} candidates escaped previous detection because of their low surface brightness, which indicates an advanced age. They are most likely evolved and expanding in a rarefied environment, and we note that they occupy the bottom part of the \ac{SNR} $\Sigma$--D diagram \citep{2018ApJ...852...84P,2020NatAs...4..910U}. In fact, the arithmetic average of surface brightness ($\Sigma_{\rm 1\,GHz}$; Table~\ref{tab:snrtable}, Column~8) from the sample of 40 \ac{LMC} \acp{SNR} \citep{2017ApJS..230....2B} is 7.9$\times10^{-21}$~W~m$^{-2}$~Hz$^{-1}$sr$^{-1}$ while from our sample of 14 new candidates is 4.5$\times10^{-22}$~W~m$^{-2}$~Hz$^{-1}$sr$^{-1}$. This order of magnitude difference suggests that we are discovering low surface brightness \acp{SNR} in the \ac{LMC}. At the same time, our sample \ac{SNR} candidate diameters are fractionally larger but because of the large overlap with the sample of established \ac{LMC} \acp{SNR} ($D_{\rm av 14SNR}$=51.2~pc and SD=16.9 vs. $D_{\rm av 71SNR}$=44.9~pc and SD=24.9) we can not separate two samples. 

\begin{figure}
\begin{center}
        \includegraphics[width=\columnwidth,trim = 20 0 30 0,clip]{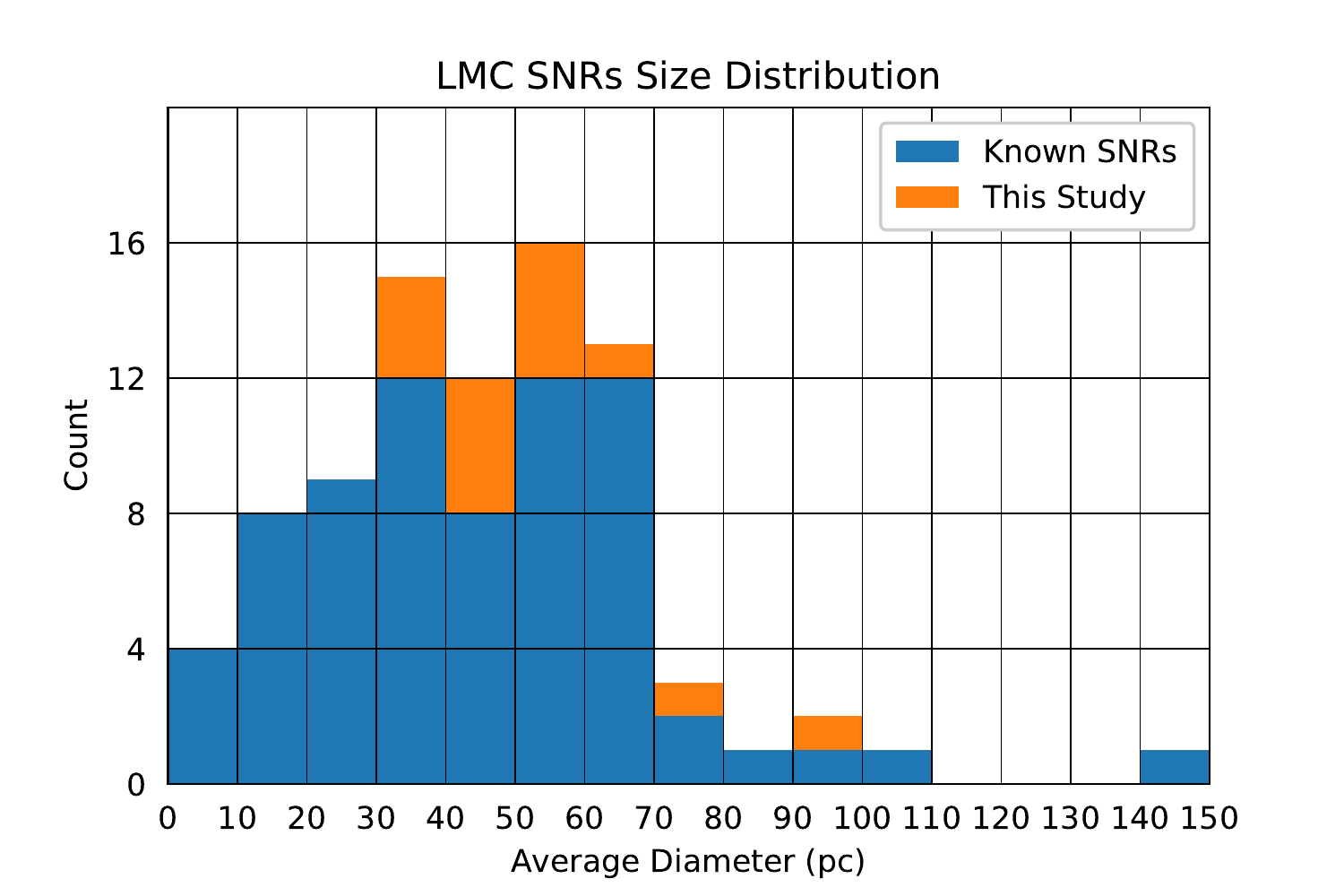}
    \caption{The distribution of 14 new \ac{LMC} \ac{SNR} candidate diameters (orange) added to the 71 previously known (blue). These 71 confirmed \acp{SNR} are detected in \citep[][]{2017ApJS..230....2B,2016A&A...585A.162M,2019MNRAS.490.5494M,2021MNRAS.504..326M,2021MNRAS.500.2336Y,2022MNRAS.tmp..786K}. }
     \label{fig:newdist}
  \end{center}
\end{figure}

The size distribution of the 14 new \ac{SNR} candidates against the 71 (58+1+1+1+3+7) previously confirmed \acp{SNR} from \citet{2017ApJS..230....2B}, \citet{2016A&A...585A.162M,2019MNRAS.490.5494M,2021MNRAS.504..326M}, \citet{2021MNRAS.500.2336Y} and \citet{2022MNRAS.tmp..786K} can be seen in Fig.~\ref{fig:newdist}. Our new sample of 14 \ac{LMC} \ac{SNR} candidates are distributed across a large range of sizes from $\sim$30 to $\sim$96~pc.  From the sample of 71 known \ac{LMC} \acp{SNR}, 45 objects have an estimated age, which spreads from the small size SN1987A of 35~yrs to the large 107~pc DEM\,L72 of 115000~yrs \citep{2010ApJ...725.2281K}. \citet[][see their fig.~18]{2017ApJS..230....2B} showed that at age of $\sim$5000~yrs \acp{SNR} such as N\,49 \citep{2012ApJ...748..117P} or 30~Dor~B \citep[a.k.a. N\,157B;][]{2006ApJ...651..237C} can reach sizes of $>$30~pc in diameter which is mid-to-late Sedov phase. However, this is very much dependent on various factors such as the environmental density, initial progenitor type and its explosion energy. Nevertheless, this implies that our sample of 14 new objects belongs to a more evolved and mid-to-older ($>5000$~yrs) \ac{SNR} population.

We also cross-matched various X-ray surveys and pointed observations including \xmm\ and {\it Chandra}. However, one would not expect to detect X-ray emission from an \ac{SNR} in the very late stage of evolution in such a rarefied environment, especially at the distance of 50~kpc, as shown in the above-mentioned case of J0624--6948 \citep{2022MNRAS.512..265F}. \ac{SNR} X-ray emission depends on \ac{ISM} density squared, while synchrotron emission scales linearly with \ac{ISM} density \citep[][]{2000immm.proc..127D,2000immm.proc..179D,2022PASP..134f1001U}. This implies that X-ray emission in less dense environments will fade much quicker than predominantly non-thermal (synchrotron) radio emission. In some cases (8/14) we can detect very weak optical emission (mainly \Ha) which indicates active non-radiative shocks from the late Sedov and radiative phases, implying an older \ac{SNR}.

Optical emission from \acp{SNR} typically originate from highly compressed, thin, radiative shells. This may further indicate the eight optically-identified \ac{SNR} candidates are not expanding in a low-density medium. In 6 candidates, we did not detect optical emission, which suggests that these \ac{SNR} candidates are in the radiative phase of their evolution. However, in the radiative phase, there could be significant emission from various coolants, depending on temperature ($T$). Most notably, we may see \OIII\ emission from $T\sim$10$^{5}$\,K gas, as the X-ray emission of such shocks fades. However, for $T\sim$10$^{5}$\,K there is little line emission in the \ac{MCELS} optical bands that we use here. This all implies that in many cases, the X-rays from \acp{SNR} fade first followed by the optical emission and then the last standing emission before the remnant completely disappears would come from radio continuum \citep{2020pesr.book.....V}. That suggests, the eight optically-detected candidates are perhaps not as old as those in which we detect only radio emission. We also detect one \ac{SNR} candidate (J0542--6852) in the \xmm\ survey which strengthens our \ac{SNR} classification for this source.

We investigate the \HI\ properties of our \ac{SNR} candidate sample (see Appendix~\ref{sec:snrs}). As a result, $\sim$80\% (11 out of 14) of them show possible evidence of expanding \HI\ shells, which were likely formed by a combination of shock waves and strong stellar winds from the massive progenitor (and possibly surrounding OB stars). As shown in the Appendix~\ref{sec:snrs}, several \acp{SNR} show a good spatial correspondence between the intensity peaks of the radio continuum and \HI, suggesting that shock-cloud interactions occurred. Indeed, our $p-v$ diagrams show cavities in which these \acp{SNR} are almost freely expanding \citep[][]{2020ApJ...902...53S,2020MNRAS.492.2606L,2019Ap&SS.364..204A}. Some of these 14 \ac{LMC} \ac{SNR} candidates (such as J0451--6906, J0457--6823, J0459--6757, J0534--6720, J0534--6700, J0542--6852 and J0543–-6923 have reached the wind-cavity of \HI, while others (such as J0452--6638, J0504--6901, J0522--6543 ) are in the free expansion phase inside the wind bubble. This is very similar to Galactic \acp{SNR} SN\,1006 \citep{2022ApJ...933..157S}, G346.6--0.2 \citep{2021ApJ...923...15S} and the mixed-morphology W49B \citep{2021ApJ...919..123S} where a wind-blown bubble was found along the radio continuum shell with an expansion velocity of $\sim$10~\kms, which was likely formed by strong stellar winds from the high-mass progenitor of the \ac{SNR}.

\begin{figure*}
  \begin{center}
    \resizebox{0.24\linewidth}{!}{{{~~~~~~~~~~~J0451-6906~~~~~~~~~~~}}}
    \resizebox{0.24\linewidth}{!}{{{~~~~~~~~~~~J0451-6951~~~~~~~~~~~}}}
    \resizebox{0.24\linewidth}{!}{{{~~~~~~~~~~~J0452-6638~~~~~~~~~~~}}}\vspace{1mm}
    \resizebox{0.24\linewidth}{!}{{{~~~~~~~~~~~J0457-6823~~~~~~~~~~~}}}
  
   \resizebox{0.24\linewidth}{!}{{\includegraphics[trim = 142 20 169 40,clip]{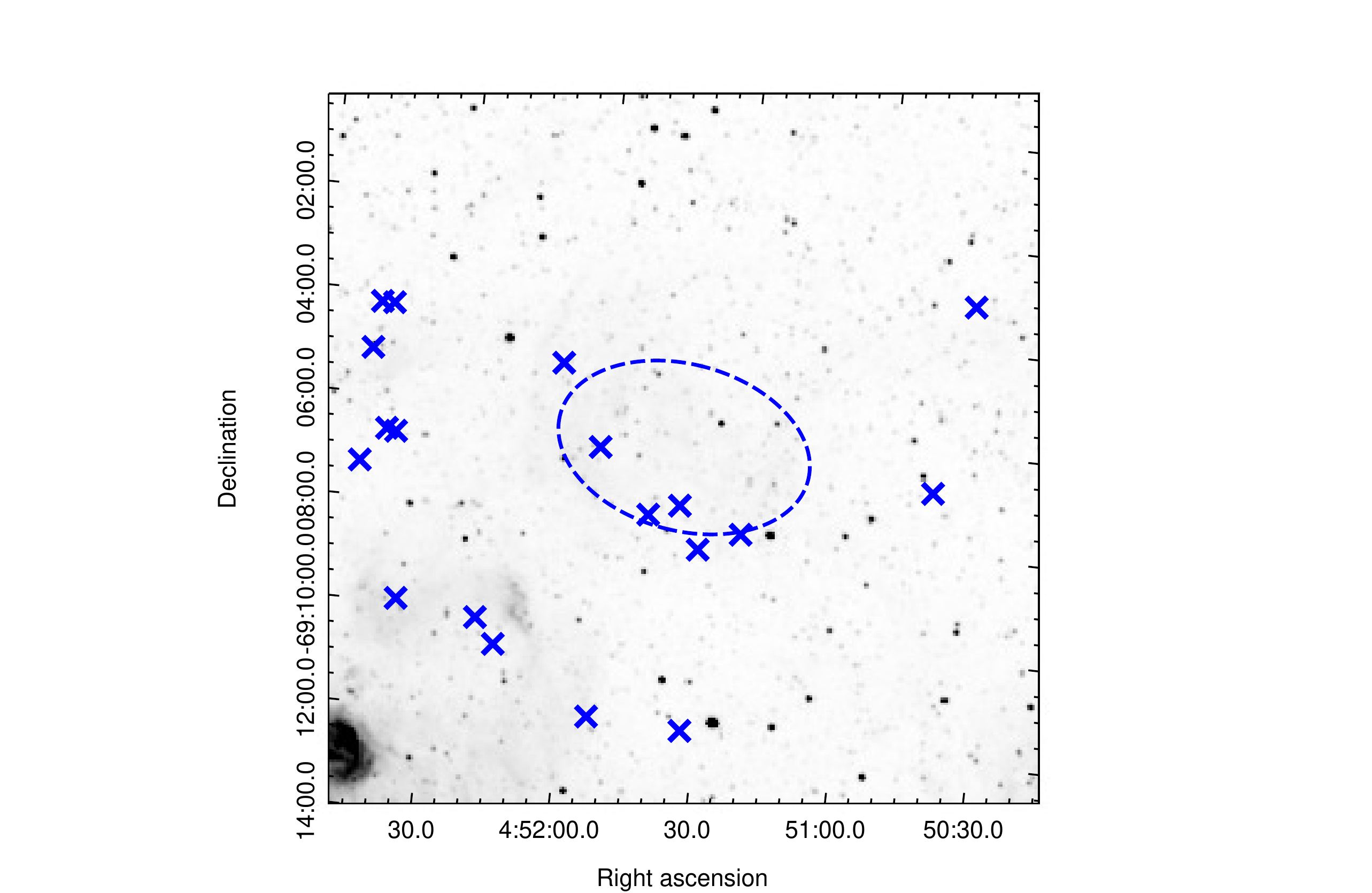}}}
   \resizebox{0.24\linewidth}{!}{{\includegraphics[trim = 142 20 169 40,clip]{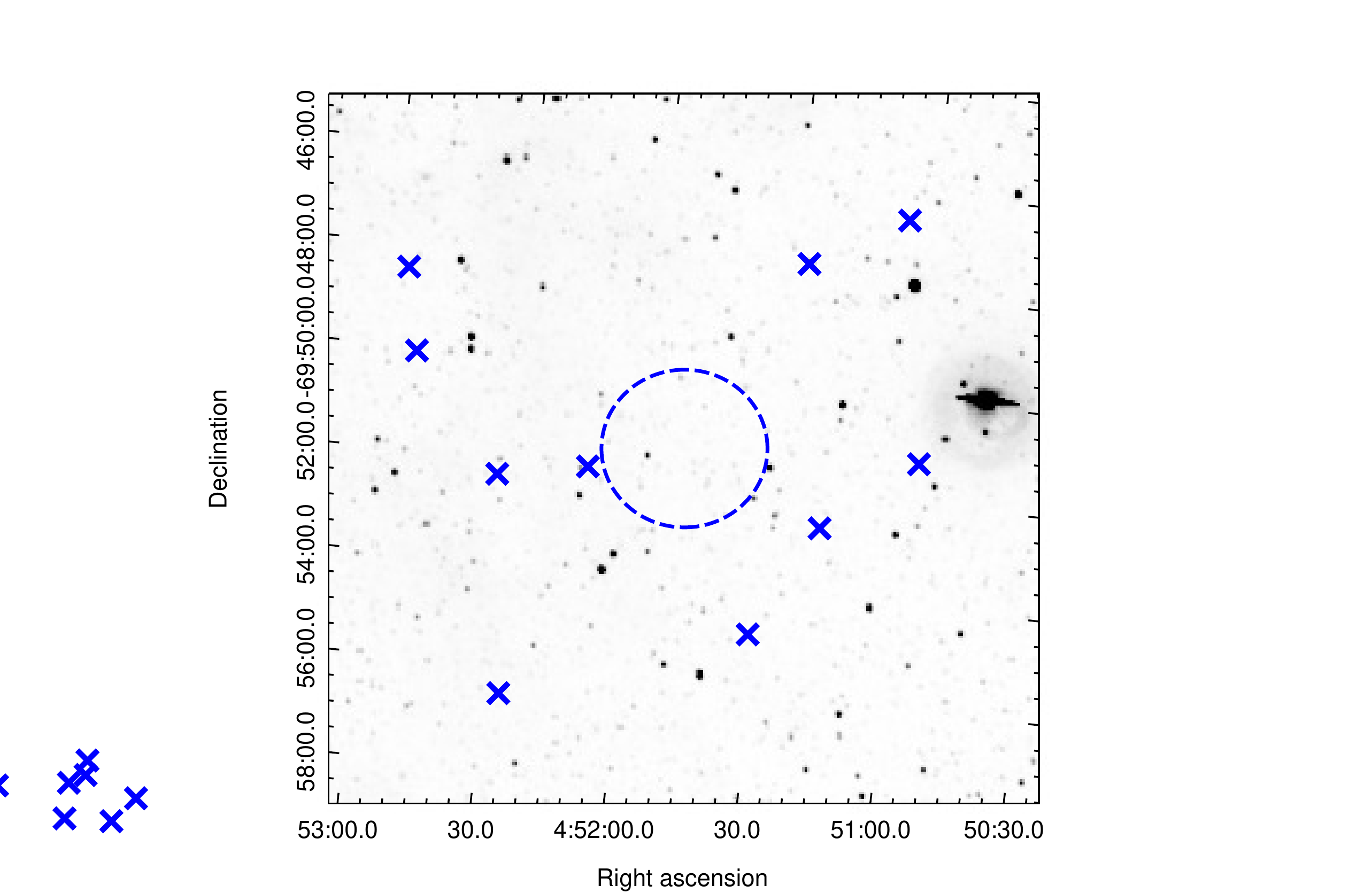}}}
   \resizebox{0.24\linewidth}{!}{{\includegraphics[trim = 142 20 169 40,clip]{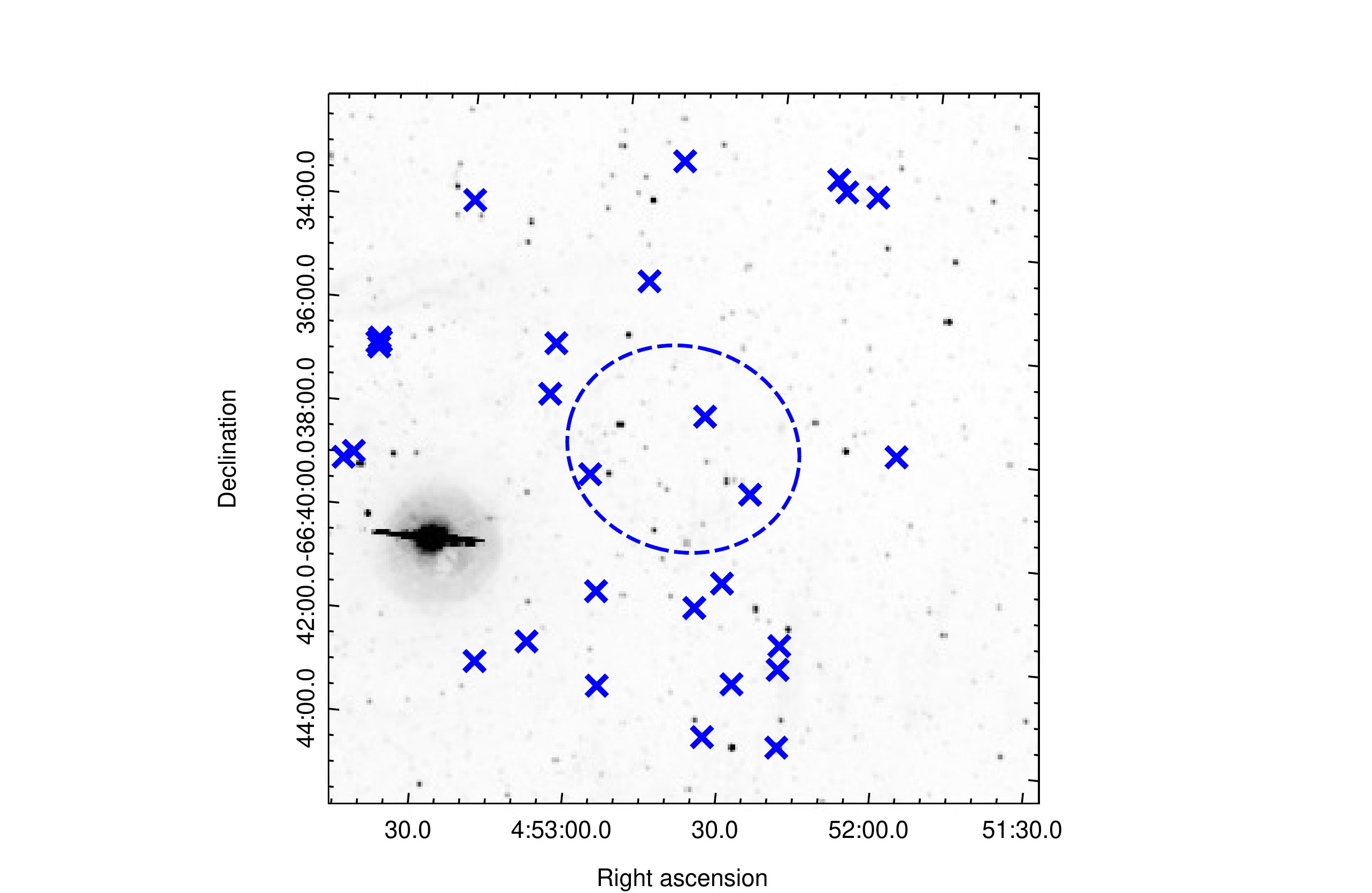}}}\vspace{5mm}
   \resizebox{0.24\linewidth}{!}{{\includegraphics[trim = 142 20 169 40,clip]{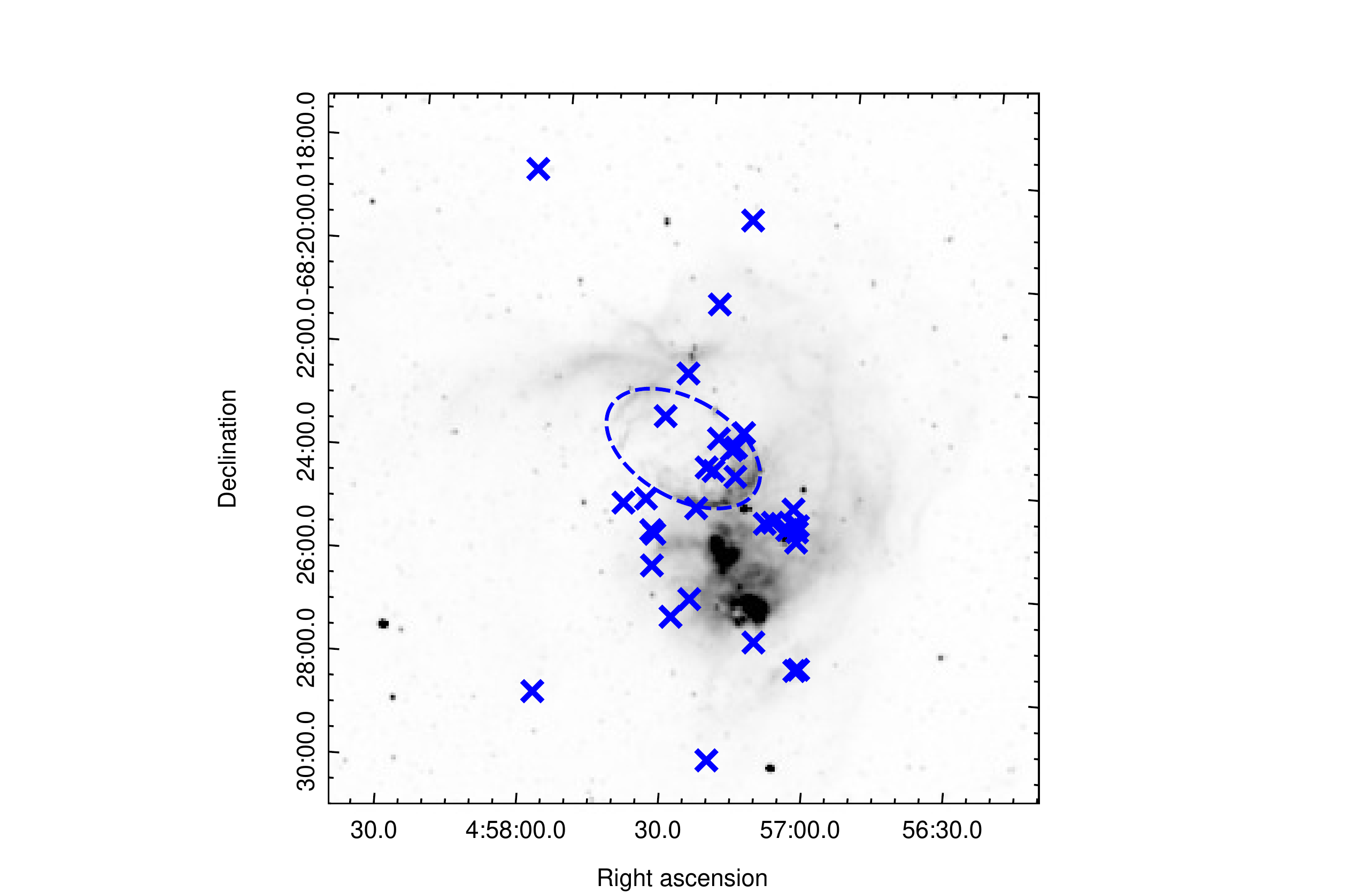}}}
   
    \resizebox{0.24\linewidth}{!}{{{~~~~~~~~~~~J0459-7008b~~~~~~~~~~~}}}
    \resizebox{0.24\linewidth}{!}{{{~~~~~~~~~~~J0459-6757~~~~~~~~~~~}}}
    \resizebox{0.24\linewidth}{!}{{{~~~~~~~~~~~J0504-6901~~~~~~~~~~~}}}\vspace{1mm}
    \resizebox{0.24\linewidth}{!}{{{~~~~~~~~~~~J0522-6543~~~~~~~~~~~}}}
    
   \resizebox{0.24\linewidth}{!}{{\includegraphics[trim = 142 20 169 40,clip]{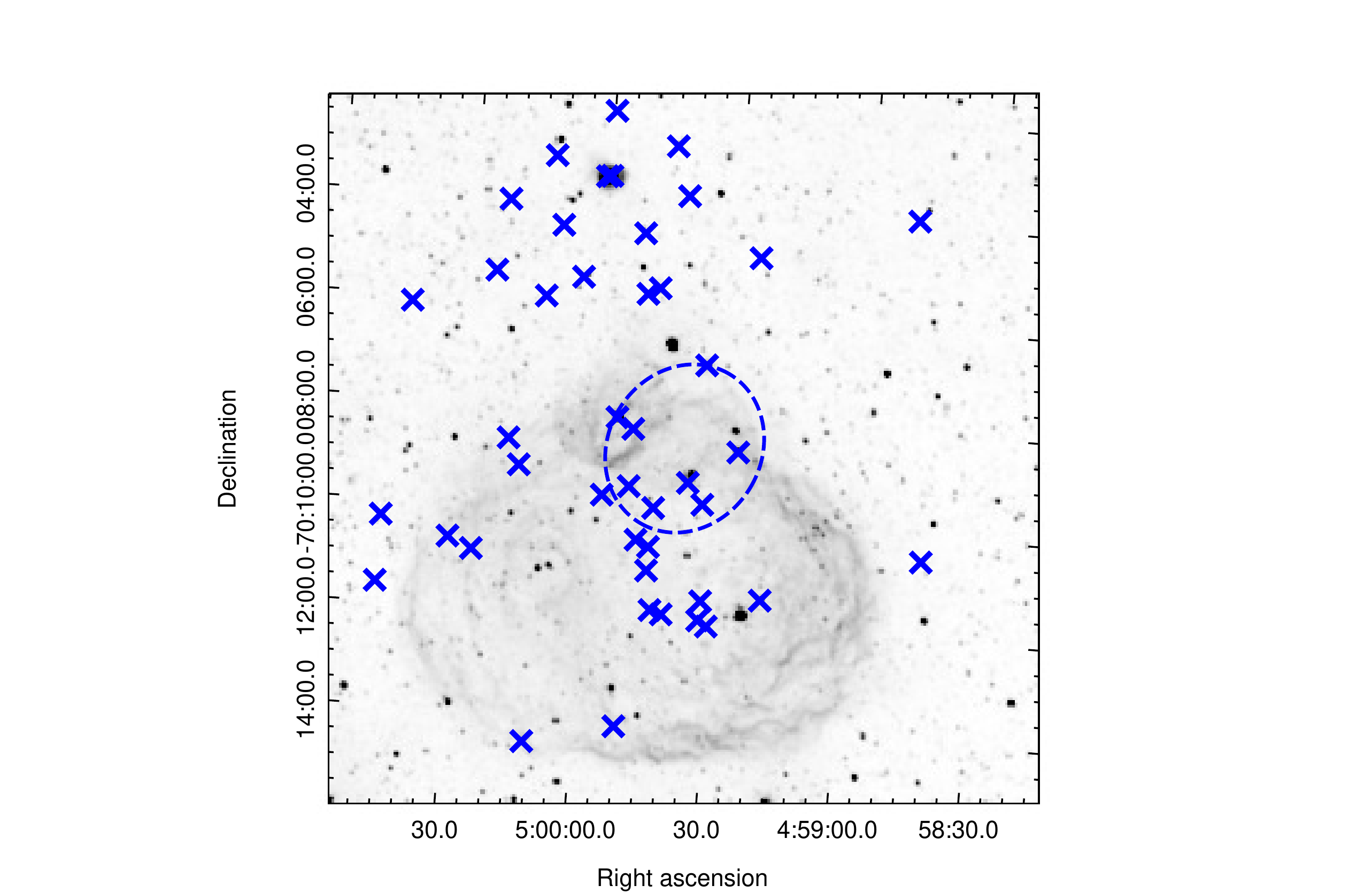}}}
   \resizebox{0.24\linewidth}{!}{{\includegraphics[trim = 142 20 169 40,clip]{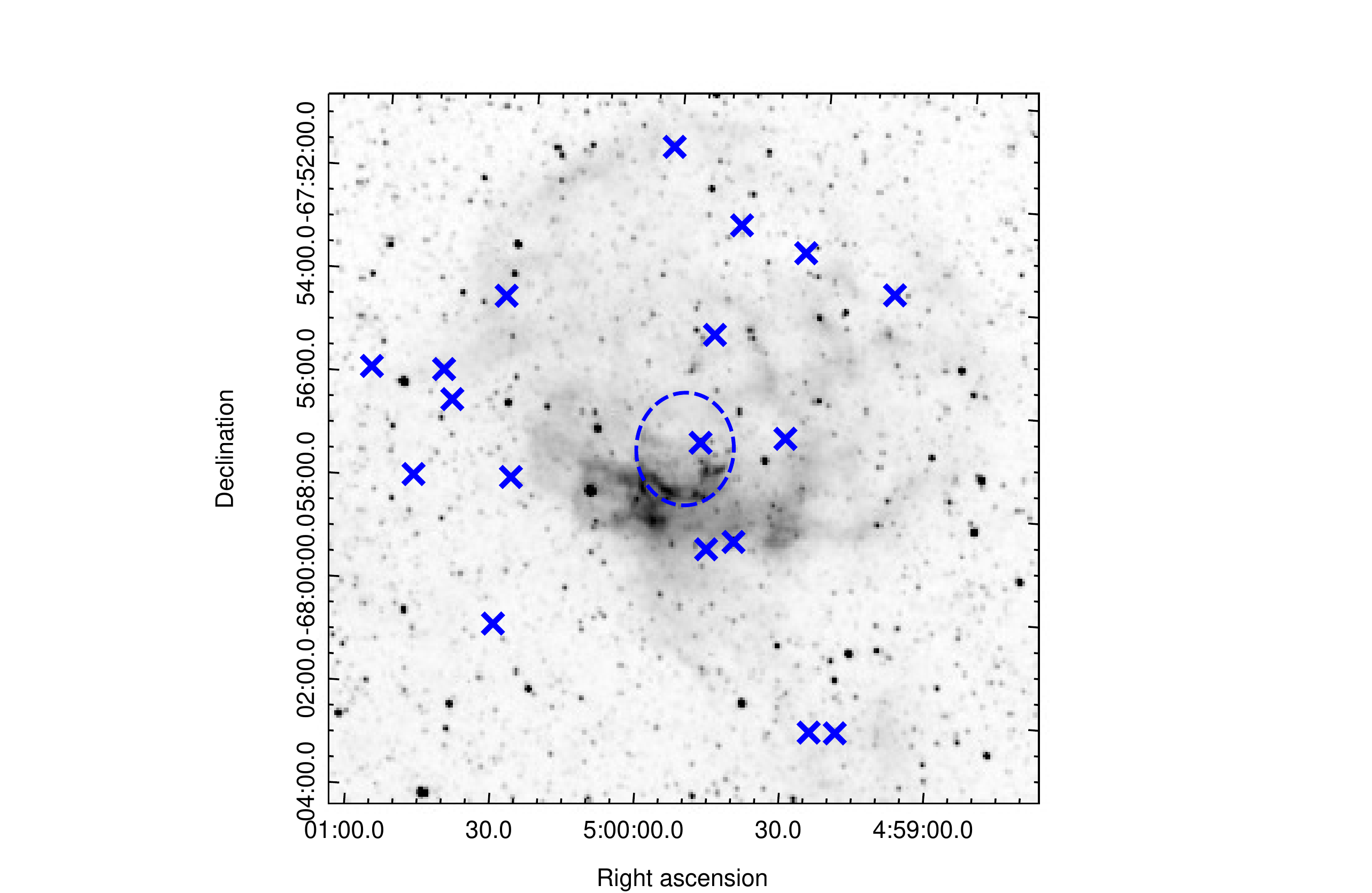}}}
   \resizebox{0.24\linewidth}{!}{{\includegraphics[trim = 142 20 169 40,clip]{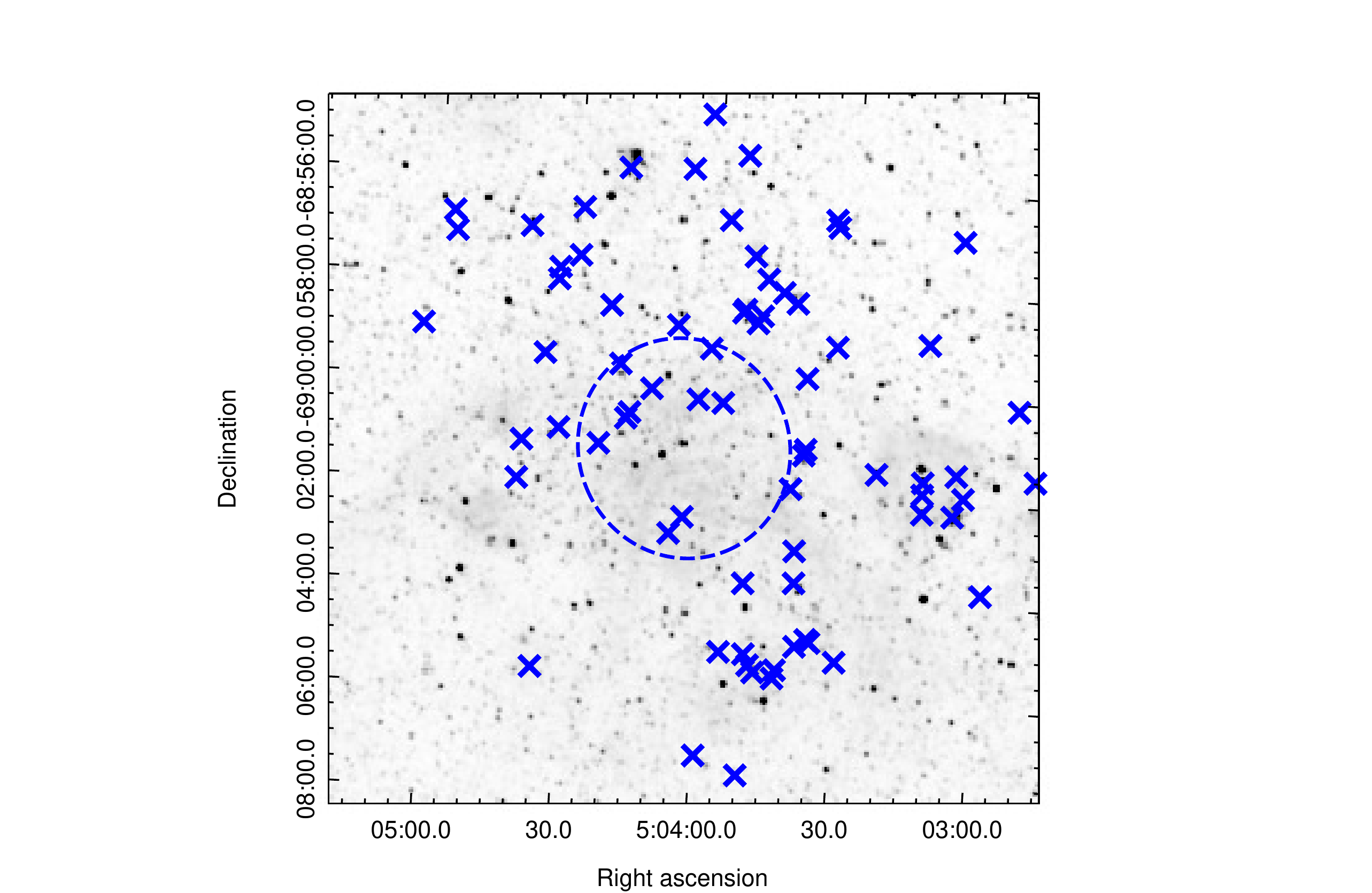}}}\vspace{5mm}
   \resizebox{0.24\linewidth}{!}{{\includegraphics[trim = 142 20 169 40,clip]{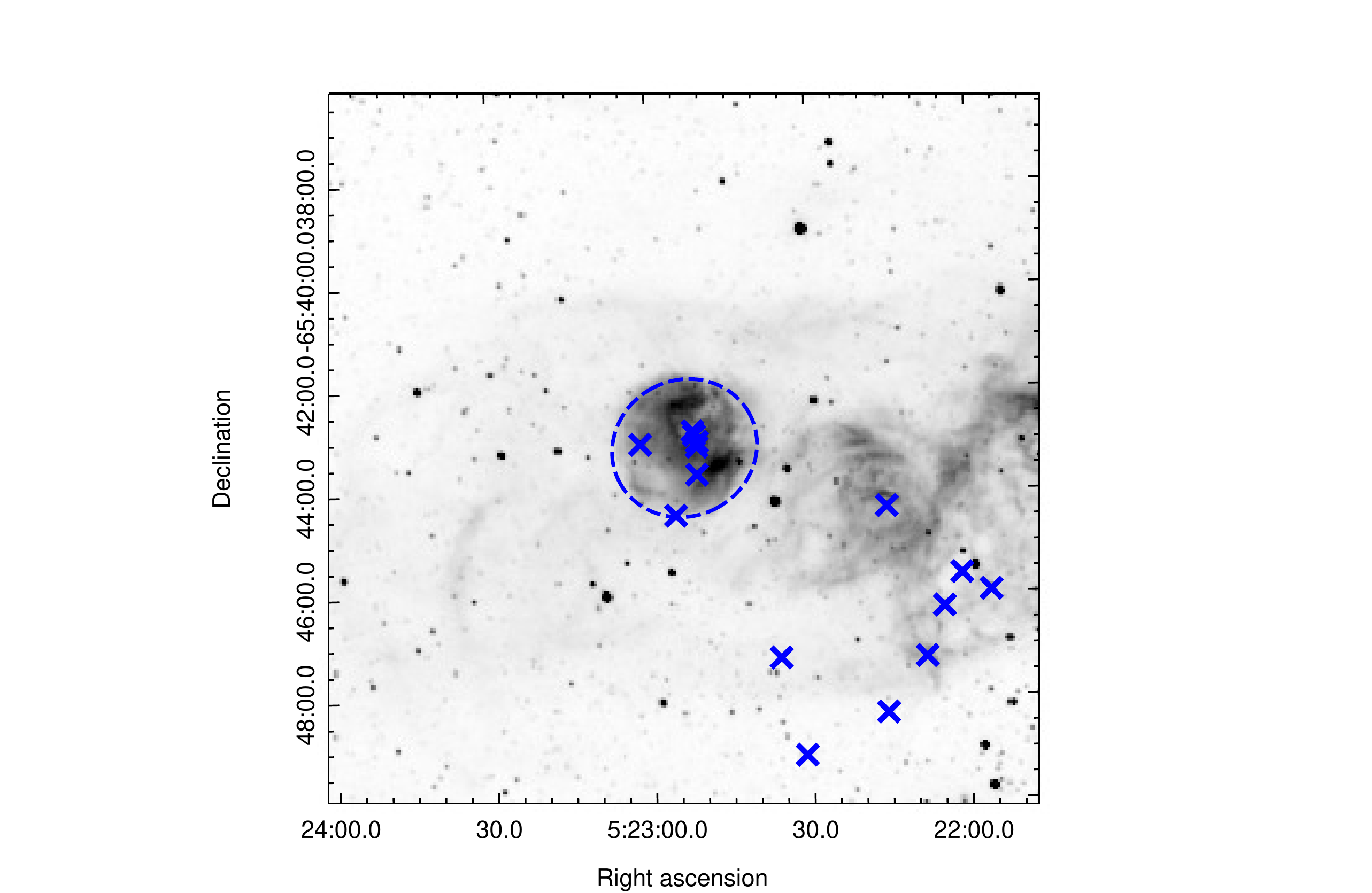}}}
   
    \resizebox{0.24\linewidth}{!}{{{~~~~~~~~~~~J0534-6720~~~~~~~~~~~}}}
    \resizebox{0.24\linewidth}{!}{{{~~~~~~~~~~~J0534-6700~~~~~~~~~~~}}}
    \resizebox{0.24\linewidth}{!}{{{~~~~~~~~~~~J0542-6852~~~~~~~~~~~}}}\vspace{1mm}
    \resizebox{0.24\linewidth}{!}{{{~~~~~~~~~~~J0543-6928~~~~~~~~~~~}}}

   \resizebox{0.24\linewidth}{!}{{\includegraphics[trim = 142 20 169 40,clip]{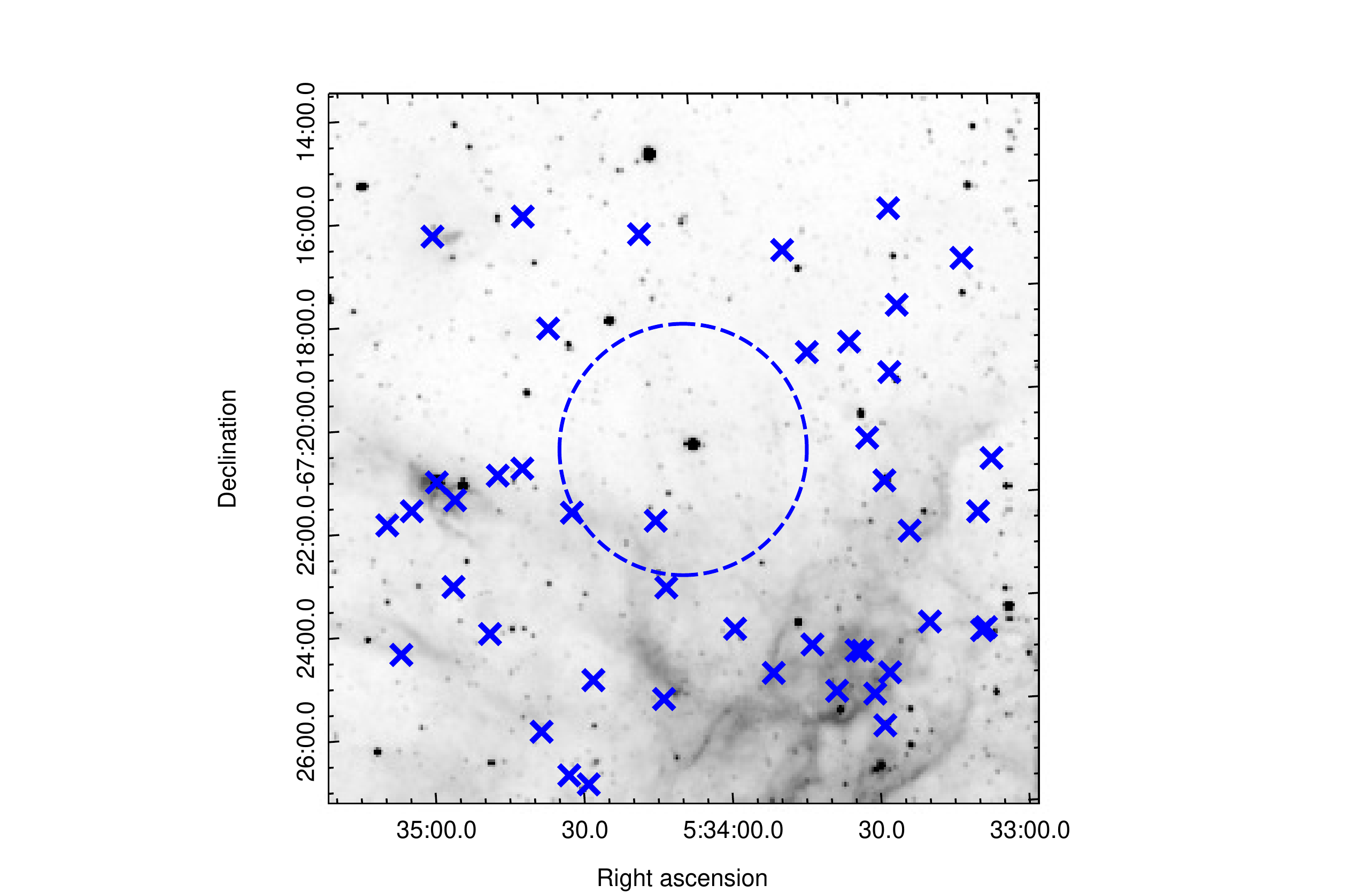}}}
   \resizebox{0.24\linewidth}{!}{{\includegraphics[trim = 142 20 169 40,clip]{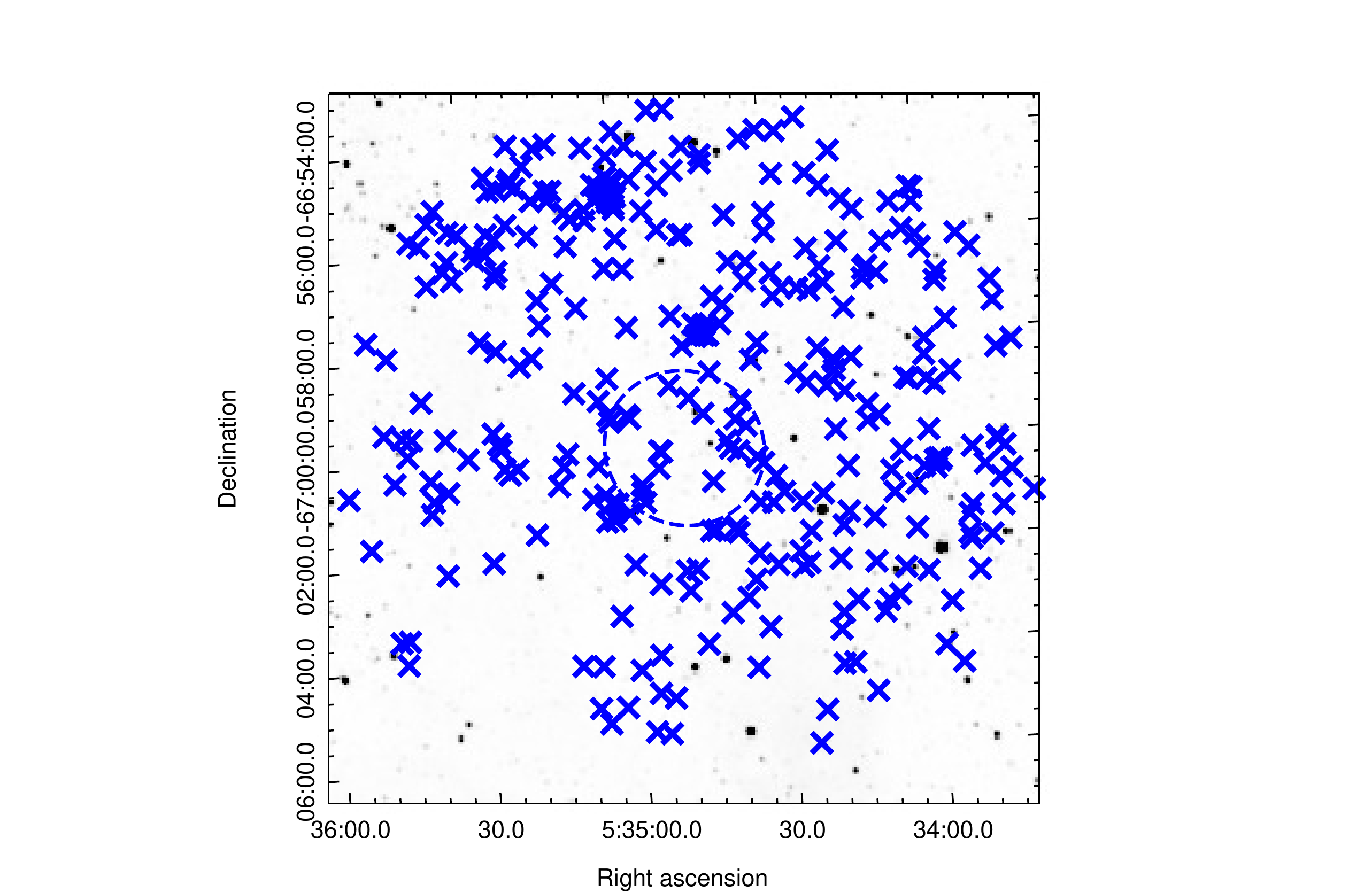}}}
   \resizebox{0.24\linewidth}{!}{{\includegraphics[trim = 142 20 169 40,clip]{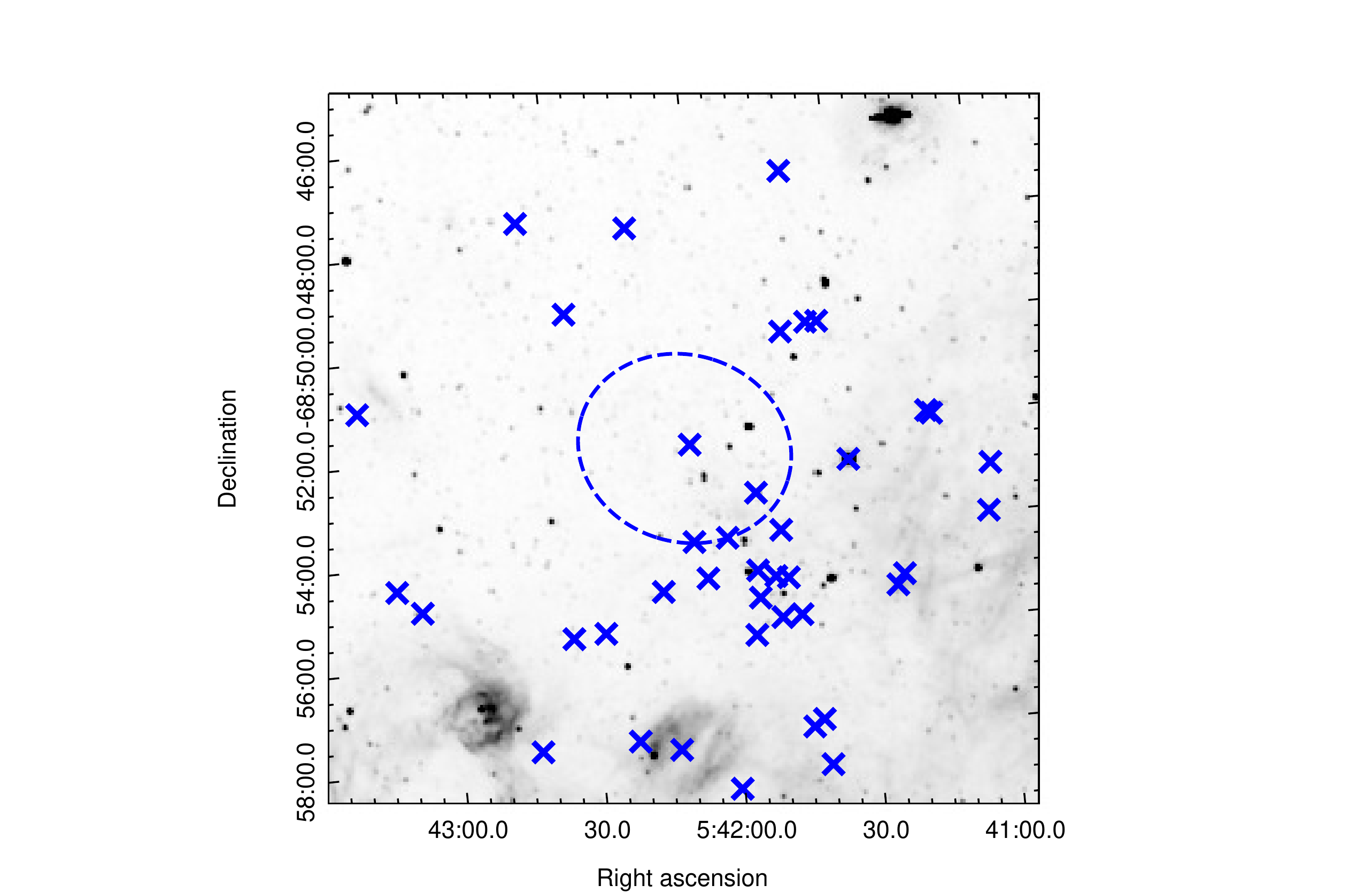}}}\vspace{5mm}
   \resizebox{0.24\linewidth}{!}{{\includegraphics[trim = 142 20 169 40,clip]{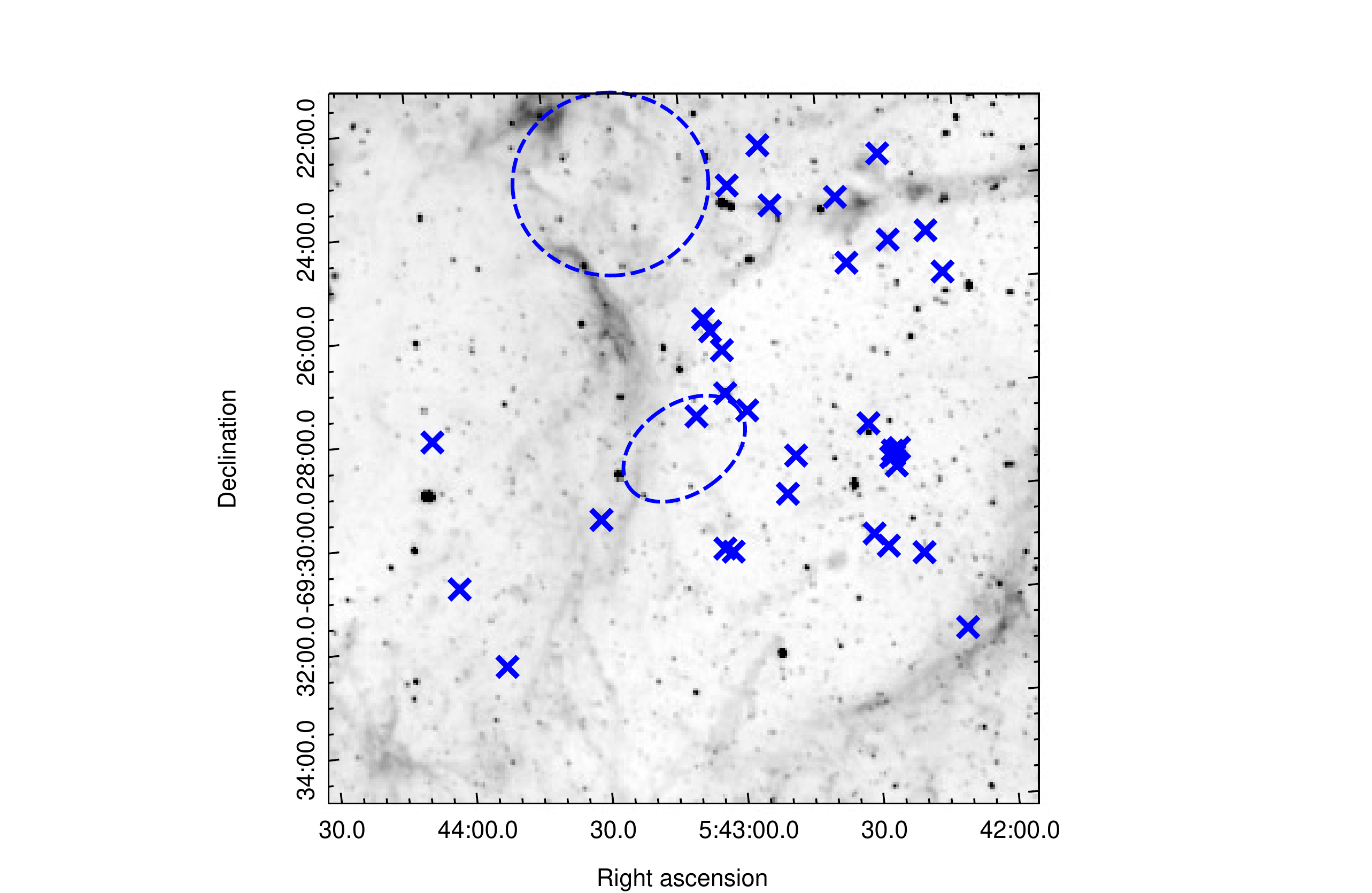}}}
   
    \resizebox{0.24\linewidth}{!}{{{~~~~~~~~~~~J0543-6923~~~~~~~~~~~}}}\vspace{1mm}
    \resizebox{0.24\linewidth}{!}{{{~~~~~~~~~~~J0543-6906~~~~~~~~~~~}}}
    
   \resizebox{0.24\linewidth}{!}{{\includegraphics[trim = 142 20 169 40,clip]{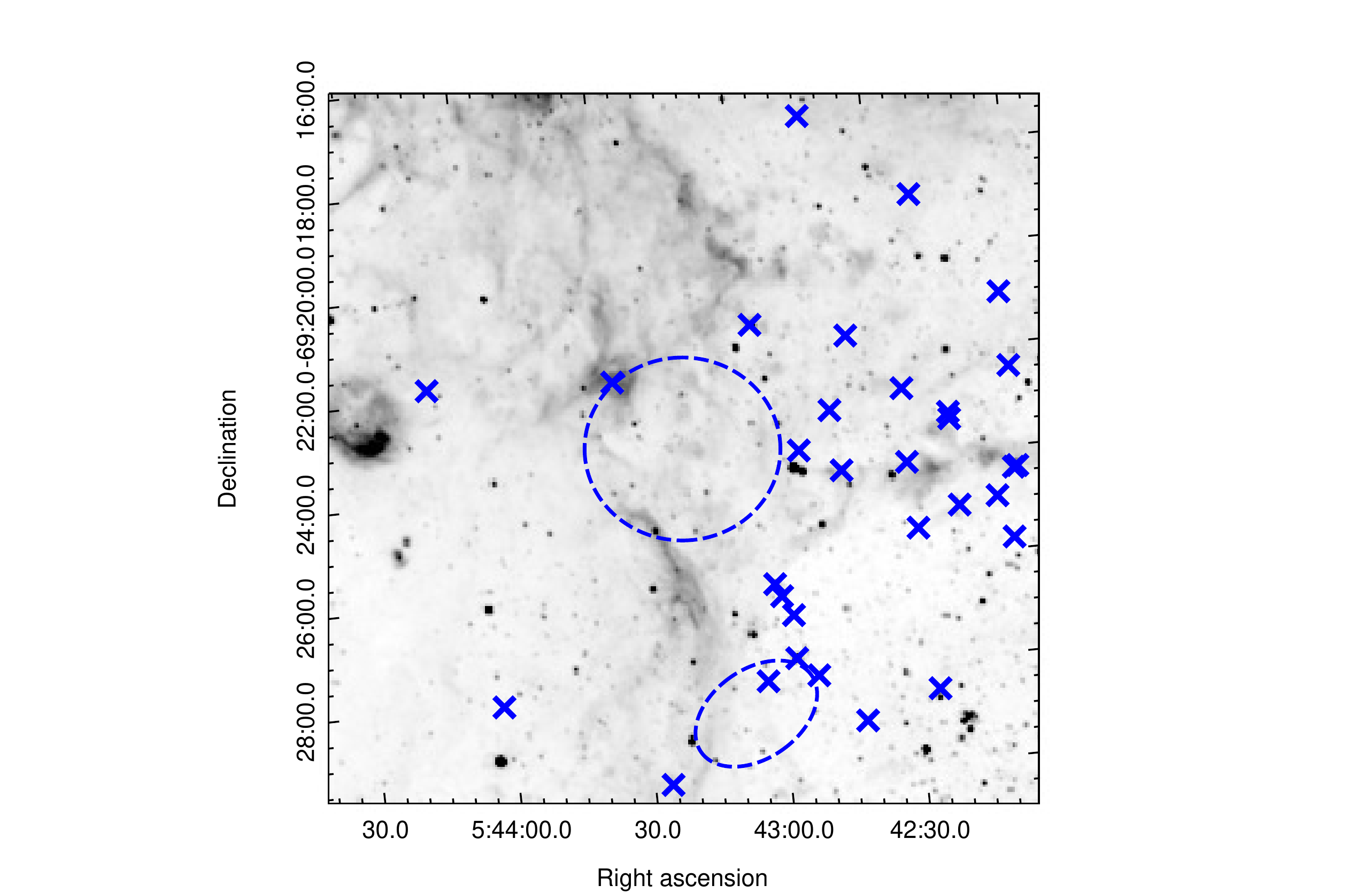}}}
   \resizebox{0.24\linewidth}{!}{{\includegraphics[trim = 142 20 169 40,clip]{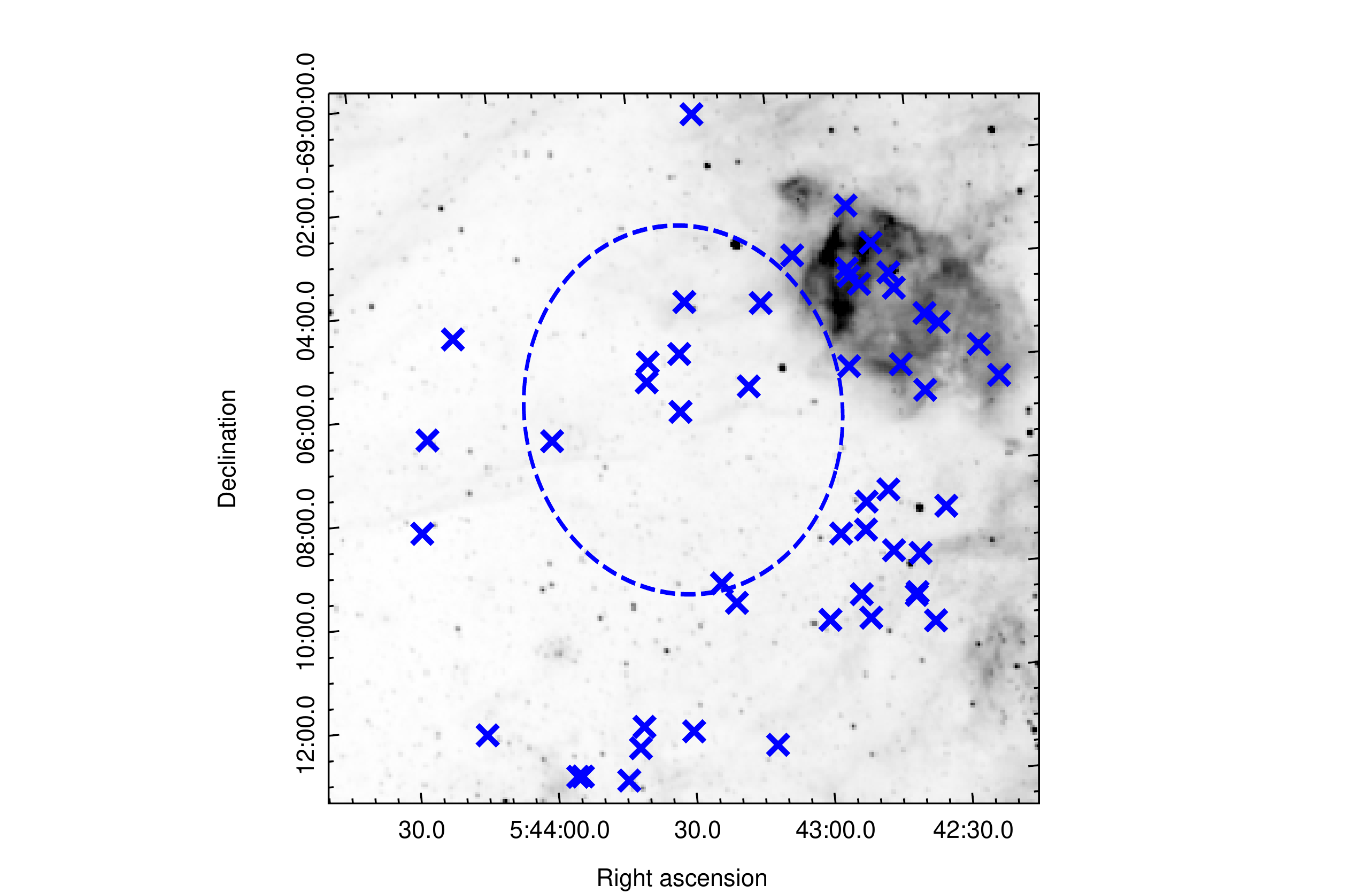}}}
   \caption{
   \ac{MCELS} images (grey-scale) of the H$\alpha$ emission around 14 new \ac{SNR} candidates in the \ac{LMC}. The dashed ellipses indicate the measured extent of the \ac{SNR} candidates (as per the data given in Table~1) and the crosses mark the positions of massive star (OB) candidates within 100~pc of the centre of the \ac{SNR} candidates. We note that J0543--6928 and J0543--6923 are sufficiently close together to appear in the each other's plot.
   }
   \label{fig:6}
  \end{center}
 \end{figure*}

Finally, we investigate the stellar environment of each of our 14 \ac{SNR} candidates (Fig.~\ref{fig:6}) using a 100~pc ($\sim$10$^{7}$~yr at 10~\kms) search radius. The number of OB stars ($N_{\rm OB}$) in the environments of these objects is an indicator of whether they are more likely to be core-collapse rather than type Ia SNRs, as such short-lived stars are a direct tracers of star formation activity that must have occurred recently in the core-collapse scenario. This indicator was calibrated in the LMC in \citet{2016A&A...585A.162M} using SNR types determined via other methods. Numbers of neighbouring OB stars less than 15 are associated with type~Ia \acp{SNR} while CC \acp{SNR} had generally (much) more than 35 such stars around them. The overall number of OB stars in our candidates trend toward the lower end of what is observed in other confirmed \ac{LMC} \acp{SNR} with several in the `undecided' range. The outlier here is J0534--6700 which has 325 blue early-type stars in the immediate vicinity of the remnant which point toward a most likely core-collapse scenario.
However, in the SN-cavities we also expect a small number of neighbouring OB stars which suggests relatively low-mass (8-10~M$_{\odot}$) SN\,II-P progenitors. The local ratio of red supergiants to OB stars might corroborate such a hypothesis as well.

\section{Conclusions}
\label{sec:conclusion}

We have presented 14 new \ac{SNR} candidates in the \ac{LMC}, adding to the ongoing effort of compiling a comprehensive set of \ac{LMC} \acp{SNR}. For one of these 14 objects (MCSNR~J0522--6543) we confirm \ac{SNR} nature based on the elevated \SII/\Ha\ ratio and non-thermal radio spectral index (see Appendix~\ref{sec:0522notes}) as well as for one previously selected candidate (MCSNR~J0506--6815). This increases the count of  \acp{SNR} in the \ac{LMC} to 73 confirmed and 32 candidates. Our radio candidates are fainter in comparison to previously known \acp{SNR} in the \ac{LMC}; their detection made possible by a new generation of radio telescopes with improved sensitivity (such as \ac{ASKAP}). Most of our candidate remnants do not have counterparts in the X-ray domain because of the non-complete coverage and poor sensitivity, but a few have optical signatures. These \ac{SNR} candidates have a low surface brightness compared to their diameter, which we interpret as indicating an evolution in a rarefied medium. This conclusion is supported by their predominantly circular morphology. We suggest a few of these remnant candidates may have a SN type~Ia origin (fewer OB stars than commonly seen) while others may be expanding in a cavity resulting from previous winds or SNe. 
Future work using deeper radio observations (at multiple wavelengths to allow the calculation of spectral indices and confirmation of their non-thermal nature) and deeper targeted X-ray and optical observations will be essential in confirming these sources as \acp{SNR}.


\section*{Acknowledgements}

The Australian SKA Pathfinder is part of the Australia Telescope National Facility which is managed by \ac{CSIRO}. Operation of \ac{ASKAP} is funded by the Australian Government with support from the National Collaborative Research Infrastructure Strategy. \ac{ASKAP} uses the resources of the Pawsey Supercomputing Centre. 
Establishment of \ac{ASKAP}, the Murchison Radio-astronomy Observatory and the Pawsey Supercomputing Centre are initiatives of the Australian Government, with support from the Government of Western Australia and the Science and Industry Endowment Fund. 
We acknowledge the Wajarri Yamatji people as the traditional owners of the Observatory site. 
This work was supported by resources provided by the Pawsey Supercomputing Centre with funding from the Australian Government and the Government of Western Australia. 
D.U. acknowledges the Ministry of Education, Science and Technological Development of the Republic of Serbia through  contract No. 451-03-68/2022-14/200104, and for support through the joint project of the Serbian Academy of Sciences and Arts and Bulgarian Academy of Sciences on the detection of extragalactic \acp{SNR} and \HII\ regions. MS acknowledges support from the Deutsche Forschungsgemeinschaft through the grants SA 2131/13-1, SA 2131/14-1, and SA 2131/15-1.
We thank the anonymous referee for comments and suggestions that greatly improved our paper.


\section*{Data Availability}

The data that support the plots/images within this paper and other findings of this study are available from the corresponding author upon reasonable request. The \ac{ASKAP} data used in this article are available through the CSIRO \ac{ASKAP} Science Data Archive (CASDA) and \ac{ATCA} data via the \ac{ATOA}.




\bibliographystyle{mnras}
\bibliography{References} 

\section*{}
Please note: Oxford University Press is not responsible for the content or functionality of any supporting materials supplied by the authors. Any queries (other than missing material) should be directed to the corresponding author for the article.

\section*{}
{\it 
$^{1}$Western Sydney University, Locked Bag 1797, Penrith South DC, NSW 2751, Australia\\
$^{2}$Faculty of Engineering, Gifu University, 1-1 Yanagido, Gifu 501-1193, Japan\\
$^{3}$Dublin Institute for Advanced Studies, Astronomy \& Astrophysics Section, 31 Fitzwilliam Place, D02 XF86 Dublin 2, Ireland \\
$^{4}$Center for Data Intensive and Time Domain Astronomy, Department of Physics and Astronomy, Michigan State University, East Lansing, MI 48824, USA\\
$^{5}$Max-Planck-Institut f\"{u}r extraterrestrische Physik, Gie{\ss}enbachstra{\ss}e 1, D-85748 Garching, Germany \\
$^{6}$Australian Astronomical Optics, AAO-Macquarie, Faculty of Science and Engineering, Macquarie University, 105 Delhi Rd, North Ryde, NSW 2113, Australia \\
$^{7}$INAF -- Osservatorio Astrofisico di Catania, via Santa Sofia 78, I-95123 Catania, Italia \\
$^{8}$The Inter-University Institute for Data Intensive Astronomy (IDIA), Department of Astronomy, University of Cape Town, Rondebosch 7701, South Africa\\
$^{9}$School of Cosmic Physics, Dublin Institute for Advanced Studies, 31 Fitzwilliam Place, Dublin 2, Ireland \\
$^{10}$ATNF, CSIRO Space \& Astronomy, PO Box 76, Epping, NSW 1710, Australia \\
$^{11}$Dominion Radio Astrophysical Observatory, Herzberg Astronomy and Astrophysics, National Research Council Canada, PO Box 248, Penticton BC V2A 6J9, Canada \\
$^{12}$Department of Physics and Astronomy, University of Calgary, University of Calgary, Calgary, Alberta, T2N 1N4, Canada\\
$^{13}$Department of Physics, University of Crete, GR-70013 Heraklion, Greece; Institute of Astrophysics, FORTH, GR-71110 Heraklion, Greece\\
$^{14}$Observatoire Astronomique de Strasbourg, Universit\'e de Strasbourg, CNRS, 11 rue de l'Universit\'e, F-67000 Strasbourg, France \\
$^{15}$Lennard-Jones Laboratories, Keele University, Staffordshire ST5 5BG, UK\\
$^{16}$Cerro Tololo Inter-American Observatory/NSF's NOIRLab, Casilla 603, La Serena, Chile\\
$^{17}$Observatory Hill Waitoki, 130 Dormer Rd, Helensville RD2, Auckland 0875, New Zealand\\
$^{18}$School of Mathematical and Physical Sciences, Macquarie University, Sydney, NSW 2109, Australia\\
$^{19}$School of Physical Sciences, The University of Adelaide, Adelaide 5005, Australia\\
$^{20}$Dr. Karl Remeis Observatory, Erlangen Centre for Astroparticle Physics, Friedrich-Alexander-Universit\"{a}t Erlangen-N\"{u}rnberg, Sternwartstra{\ss}e 7, 96049 Bamberg, Germany \\
$^{21}$Department of Physics and Astronomy, University of Manitoba, Winnipeg, MB R3T 2N2, Canada\\
$^{22}$Department of Astronomy, Faculty of Mathematics, University of Belgrade, Studentski trg 16, 11000 Belgrade, Serbia\\
$^{23}$Isaac Newton Institute of Chile, Yugoslavia Branch
}



\appendix



\section{Properties of 14 newly discovered LMC SNR candidates}
\label{sec:snrs}

We present the properties of each \ac{LMC} remnant candidate detected in this study such as their extent, radio morphology, potential OB associations and counterpart emission at other wavelengths.

\subsection{J0451--6906}
\label{sec:05416906notes}

\begin{figure*}
  \begin{center}
    \resizebox{0.975\linewidth}{!}{\includegraphics[trim = 0 0 0 0,clip]{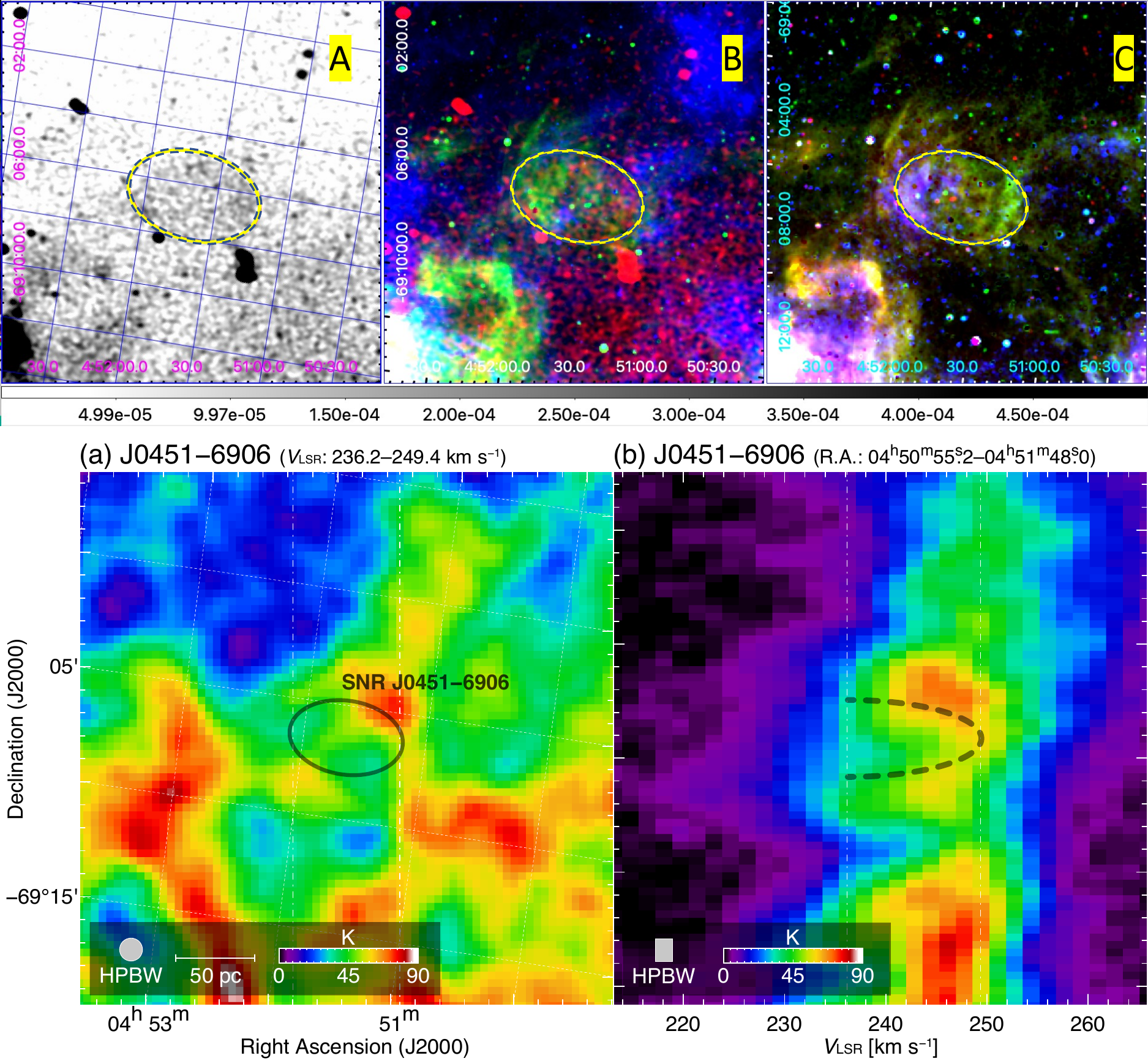}}
   \caption{J0451--6906: (Top) 
   Left (A): New \ac{ASKAP} 888~MHz radio image at a spatial resolution of 13.9$\times$12.1~arcsec$^2$ and position angle $-84\degr$. Gray scale at the bottom is from 0 to 0.5~mJy~beam$^{-1}$.
   Middle (B): RGB image where \ac{ASKAP} 888~MHz radio image is in red, \Ha\ (green) and Spitzer LMC-SAGE at 8$\mu$m in blue. 
   Right (C): \ac{MCELS} optical RGB image where \Ha\ (red), \SII\ (green), \& \OIII\ (blue). 
   (Bottom) Left (D): Integrated intensity map of H{\sc i} obtained with \ac{ATCA} \& Parkes \citep{2003ApJS..148..473K}. The vertical dash-dotted (white) lines indicate the R.A. integration range for the position--velocity diagram. The blue-yellow ellipse indicates positions of here proposed \ac{SNR}.
   Right (E): Position--velocity diagram of H{\sc i}. 
   The dashed curve delineates a cavity-like structure of \HI.
      The vertical dash-dotted lines indicate the velocity integration range for the \HI\ integrated intensity map.}
   \label{fig:0451-6906}
  \end{center}
\end{figure*}

The emission of this potential \ac{SNR} from the \ac{ASKAP} radio image appears as a faint shell morphology with slight rim brightening toward the western region (Figs.~\ref{fig:X} and~\ref{fig:0451-6906}). The western rim of this emission also has elevated \SII/\Ha\ ($>$0.5), which helps push the case for the \ac{SNR} candidate classification. Five of the 19 massive  (OB) stars close to this candidate are  positioned in the south/eastern rim of this tentative remnant. Its optical and radio boundaries are not obvious but our \HI\ and $p-v$ diagram shows evidence of a cavity-like structure typical of a \ac{SNR} expanding inside  this cavity.

\subsection{J0451--6951}
\label{sec:05416951notes}

\begin{figure*}
  \begin{center}
    \resizebox{0.975\linewidth}{!}{\includegraphics[trim = 0 0 0 0,clip]{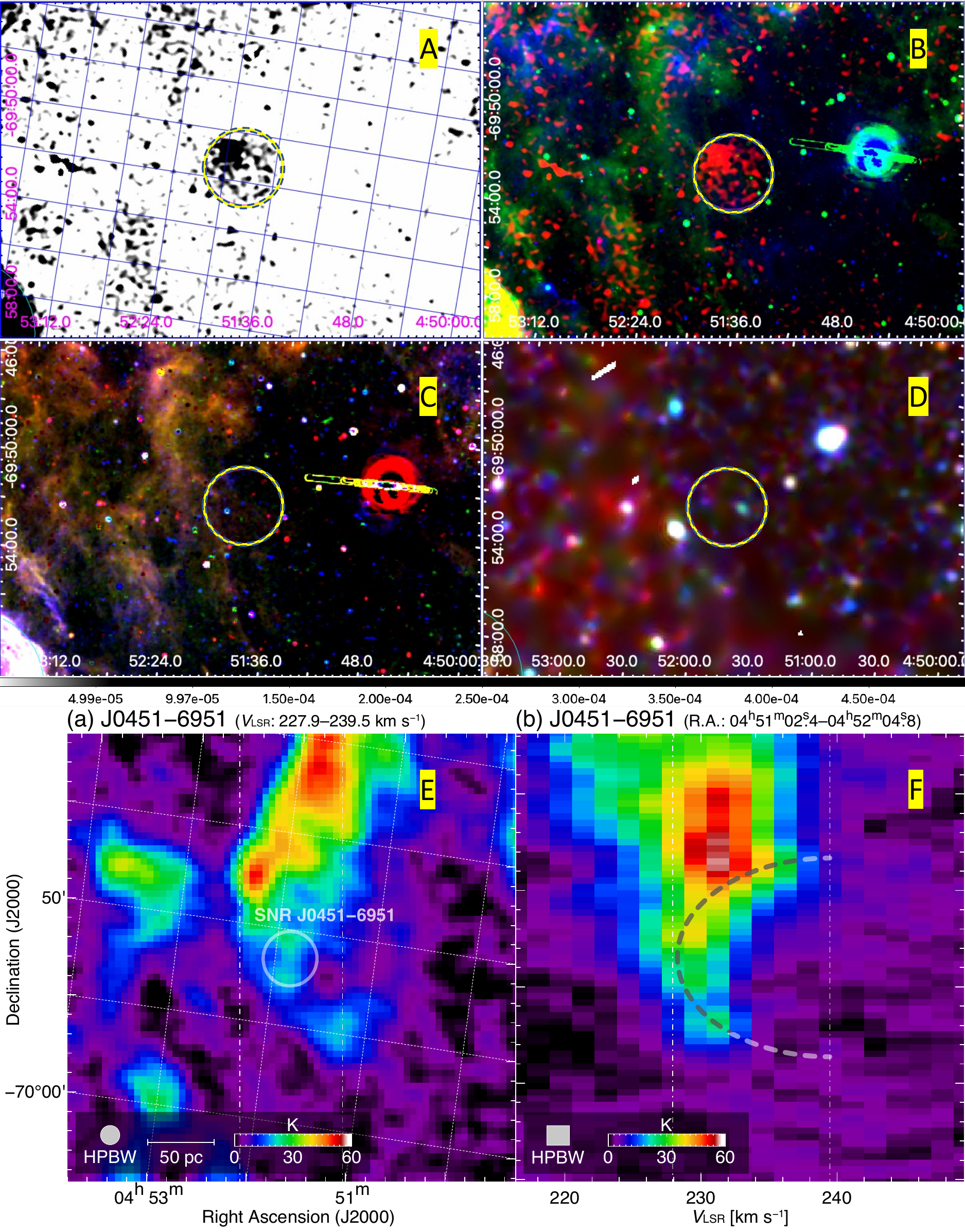}}
    \caption{J0451--6951: 
    (Top) Left (A): New \ac{ASKAP} 888~MHz radio image at spatial resolution of 13.9$\times$12.1~arcsec$^2$ and position angle $-84\degr$. Gray scale at the bottom is from 0 to 0.5~mJy~beam$^{-1}$. 
    Right (B): RGB image where \ac{ASKAP} 888~MHz radio image is in red, \Ha\ (green) and Spitzer LMC-SAGE at 8$\mu$m in blue. 
    (Middle) Left (C): \ac{MCELS} optical RGB image where \Ha\ (red), \SII\ (green), \& \OIII\ (blue). 
    Right (D): \xmm\ X-ray images using the soft (0.3–0.7~keV; red), medium (0.7–1.1~keV; green), and hard (1.1–4.2~keV; blue) bands. 
    The blue-yellow ellipse indicates positions of here proposed \ac{SNR}.
    (Bottom) Left (E): Integrated intensity map of H{\sc i} obtained with \ac{ATCA} \& Parkes \citep{2003ApJS..148..473K}.  
    Right (F): Position--velocity diagram of H{\sc i}. 
    We note image artefacts in the optical (B and C) and X-ray (D) panels which are caused by the removal of noisy pixels and by CCD chip gaps.}
   \label{fig:0451-6951}
  \end{center}
\end{figure*}

This source exhibits a circular shell morphology (Fig.~\ref{fig:X} and~\ref{fig:0451-6951}). Its diameter, $D=41$\,pc, is a typical size for the \ac{LMC} \acp{SNR} sample (where the average is 44.9~pc). 10~OB star candidates are located within 100~pc of the source, but none are located within the measured extent of the candidate. No obvious optical or X-ray emission was found.

\subsection{J0452--6638}
\label{sec:05426638notes}

\begin{figure*}
  \begin{center}
    \resizebox{0.975\linewidth}{!}{\includegraphics[trim = 0 0 0 0,clip]{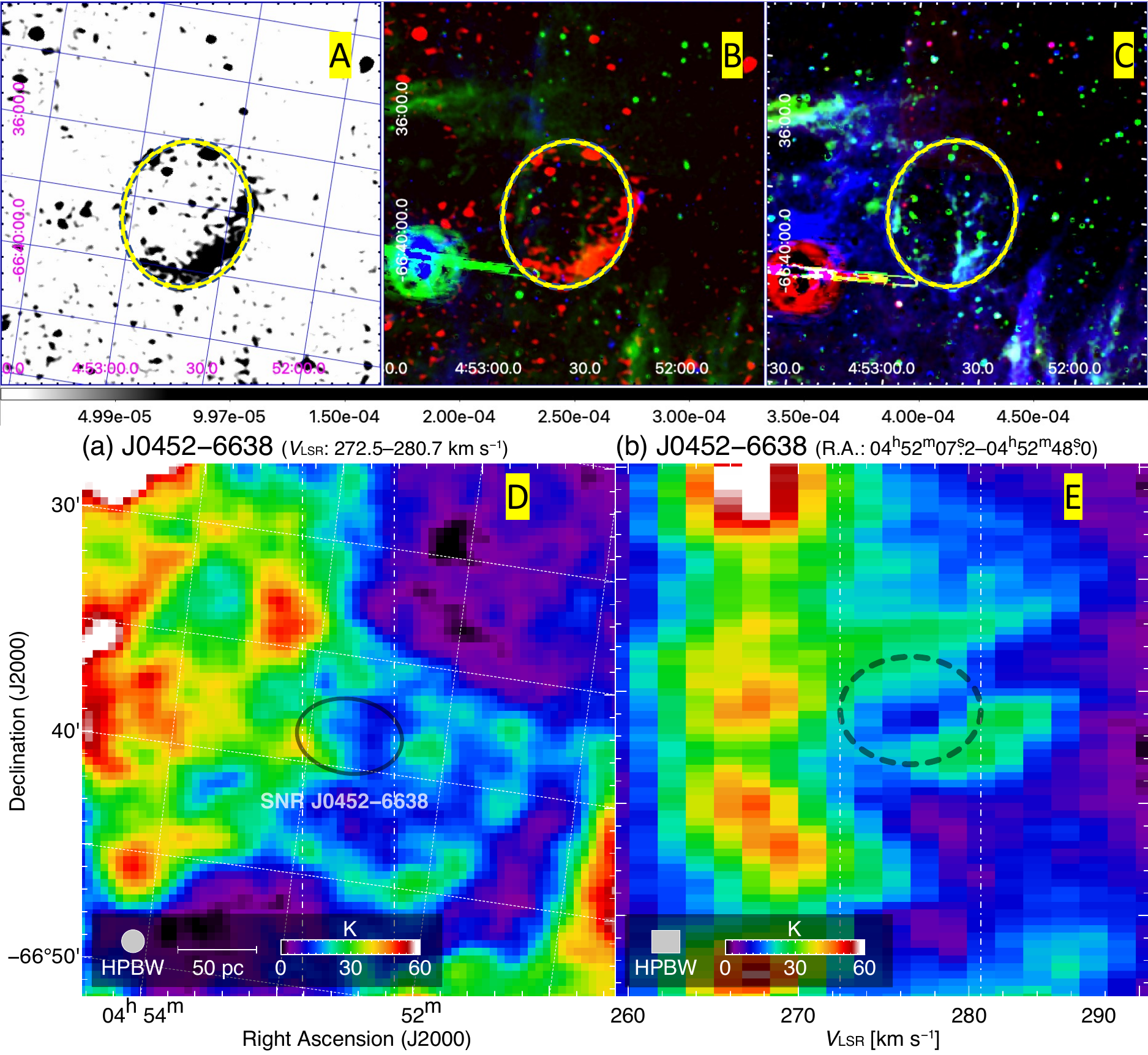}}
   \caption{J0452--6638: Same as Fig.~\ref{fig:0451-6906}. We note image artefacts in the optical (B and C) panels.}
   \label{fig:0452-6638}
  \end{center}
\end{figure*}

This \ac{SNR} candidate exhibits a shell, which is brightest to the south and west (Figs.~\ref{fig:X} and~\ref{fig:0452-6638}), with $D=56$\,pc and a very low surface brightness. Its radio emission coincides with the optical extent as it brightens on the south-west limb. There are 29 OB star candidates within 100~pc of the source, with four of these located within the extent of the ring emission. \HI\ and $p-v$ diagram shows a cavity typical for \acp{SNR} \citep[][]{2020ApJ...902...53S,2020MNRAS.492.2606L,2019Ap&SS.364..204A}.

\subsection{J0457--6823}
 \label{sec:04576823notes}

\begin{figure*}
  \begin{center}
    \resizebox{0.975\linewidth}{!}{\includegraphics[trim = 0 0 0 0,clip]{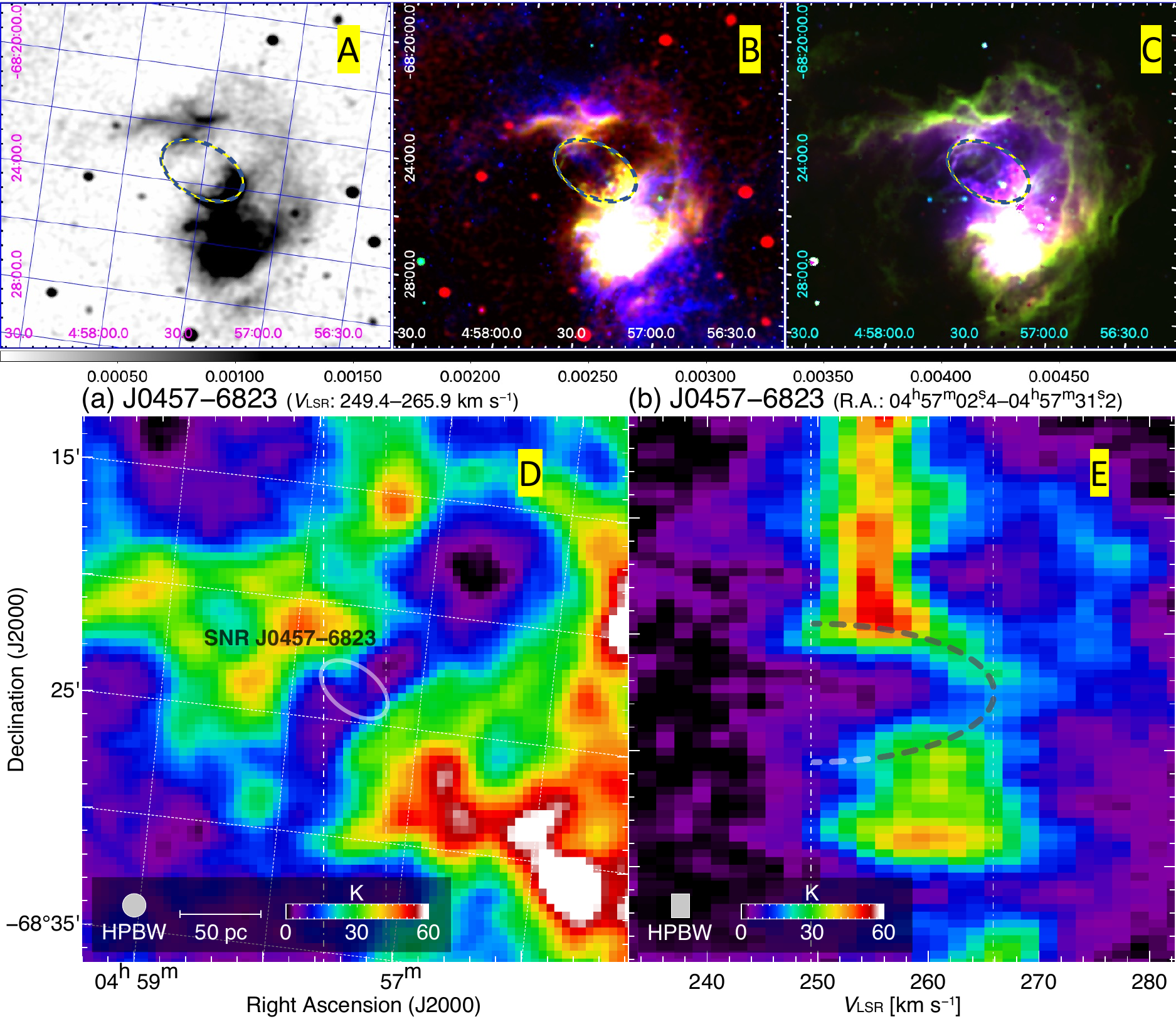}}
    \caption{J0457--6823: Same as Fig.~\ref{fig:0451-6906}. Gray scale at the bottom of panel A (\ac{ASKAP} image) is from 0 to 5~mJy~beam$^{-1}$.}   
   \label{fig:0457-6823}
  \end{center}
\end{figure*}

This \ac{SNR} candidate exhibits an elongated structure that can be seen best in the south-west where the source is possibly colliding, overlapping or embedded in emission from a separate nearby source (Figs.~\ref{fig:X} and~\ref{fig:0457-6823}). A faint extended source at this position was detected at soft X-ray energies by \citet{1999A&AS..139..277H}; source number 655 in their catalogue ([HP99]~655). The X-ray existence likelihood is low, due to the large off-axis angle of the source during the {\it ROSAT} observation and better data are needed to confirm any association with the \ac{SNR} candidate. There are 33 OB star candidates located within, or in close proximity to this source --  seven of the eight  located inside the bounds of the remnant are located in the south-western region where there is higher radio emission. The \HI\ map and $p-v$ diagram show the existence of a cavity but the \SII/\Ha\ ratio of 0.35 does not strongly favour an \ac{SNR}. Therefore, we classify this object as a low confidence \ac{SNR} candidate.


\subsection{J0459--7008b}

\begin{figure*}
  \begin{center}
    \resizebox{1\linewidth}{!}{\includegraphics[trim = 0 0 0 0,clip]{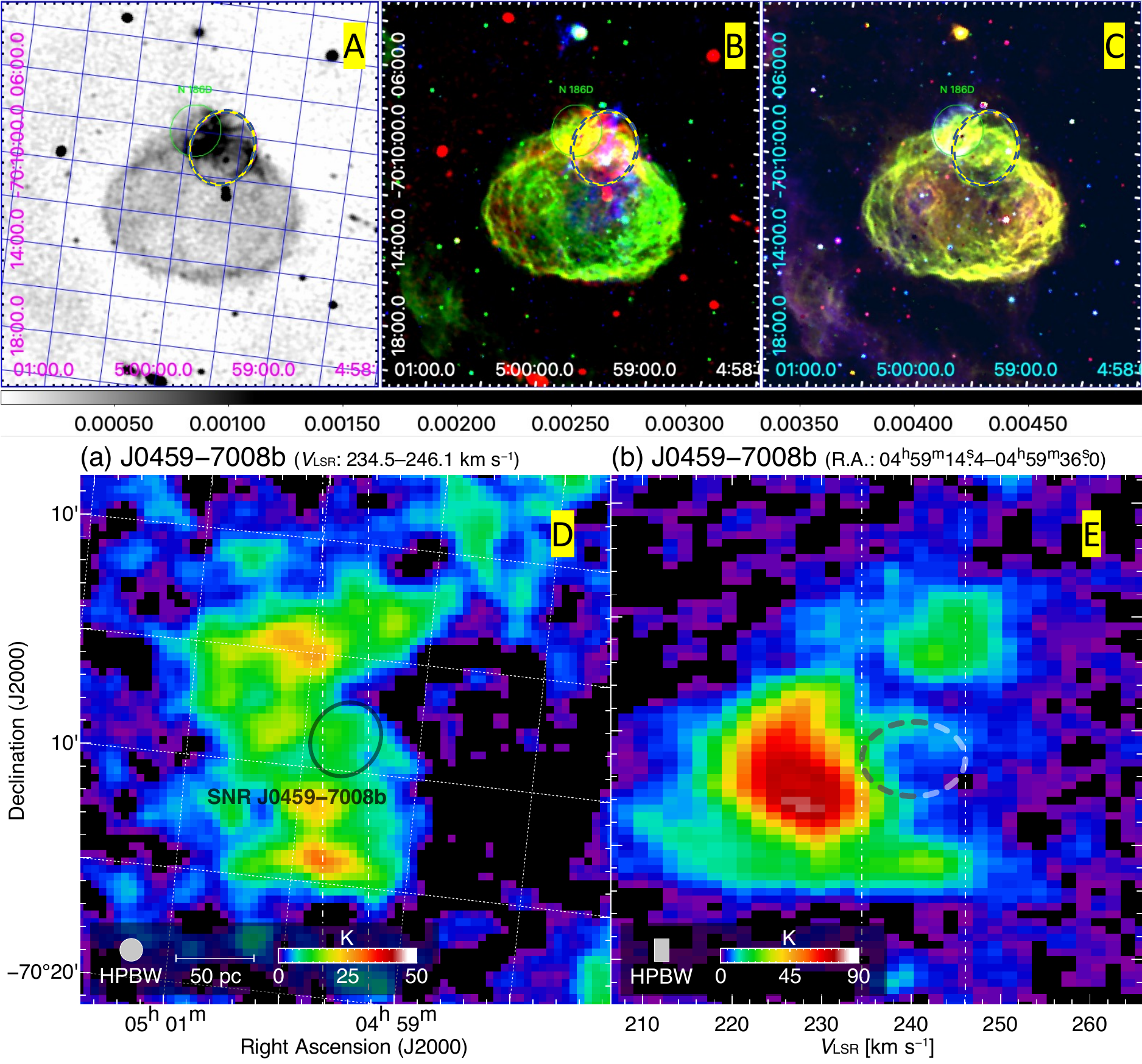}}
   \caption{J0459--7008b: Same as Fig.~\ref{fig:0451-6906}. Gray scale at the bottom of panel A (\ac{ASKAP} image) is from 0 to 5~mJy~beam$^{-1}$.}
   \label{fig:0459-7008b}
  \end{center}
\end{figure*}

The emission in this region is obfuscated by emission in the south from the massive superbubble N\,186 \citep{2002AJ....123..255O}. Located only 90~arcsec towards the east, the previously established MCSNR~J0459--7008 (a.k.a. N\,186D) with a diameter of 34.3\,pc \citep{2011ApJ...729...28J} is indicated by a bright shell in the centre left of Fig.~\ref{fig:X}. There appears to be an almost circular ring of emission ($D$=43\,pc) located to the west from the previously established \ac{SNR} (Figs.~\ref{fig:X} and~\ref{fig:0459-7008b}). In addition, there are two more faint radio shells to the north and one to the south-east. The elevated \SII/\Ha ($>$0.7) and radio-to-\Ha\ levels also have a \ac{SNR} shell-type morphology covering the same general area. In a search for massive stars in the region, we find 44 OB star candidates with $\sim6$ within the projected bounds of the region. Also, the \HI\ map and the $p-v$ diagram show a feature and possible cavity (see Fig.~\ref{fig:0459-7008b}; Panel~E). Finally, we note that \ac{SNR} N\,186D is very bright in \OIII\ while here proposed \ac{SNR} candidate J0459--7008b is bright in \SII. This is good indication that we see two different \acp{SNR}.

\subsection{J0459--6757}

\begin{figure*}
  \begin{center}
    \resizebox{0.975\linewidth}{!}{\includegraphics[trim = 0 0 0 0,clip]{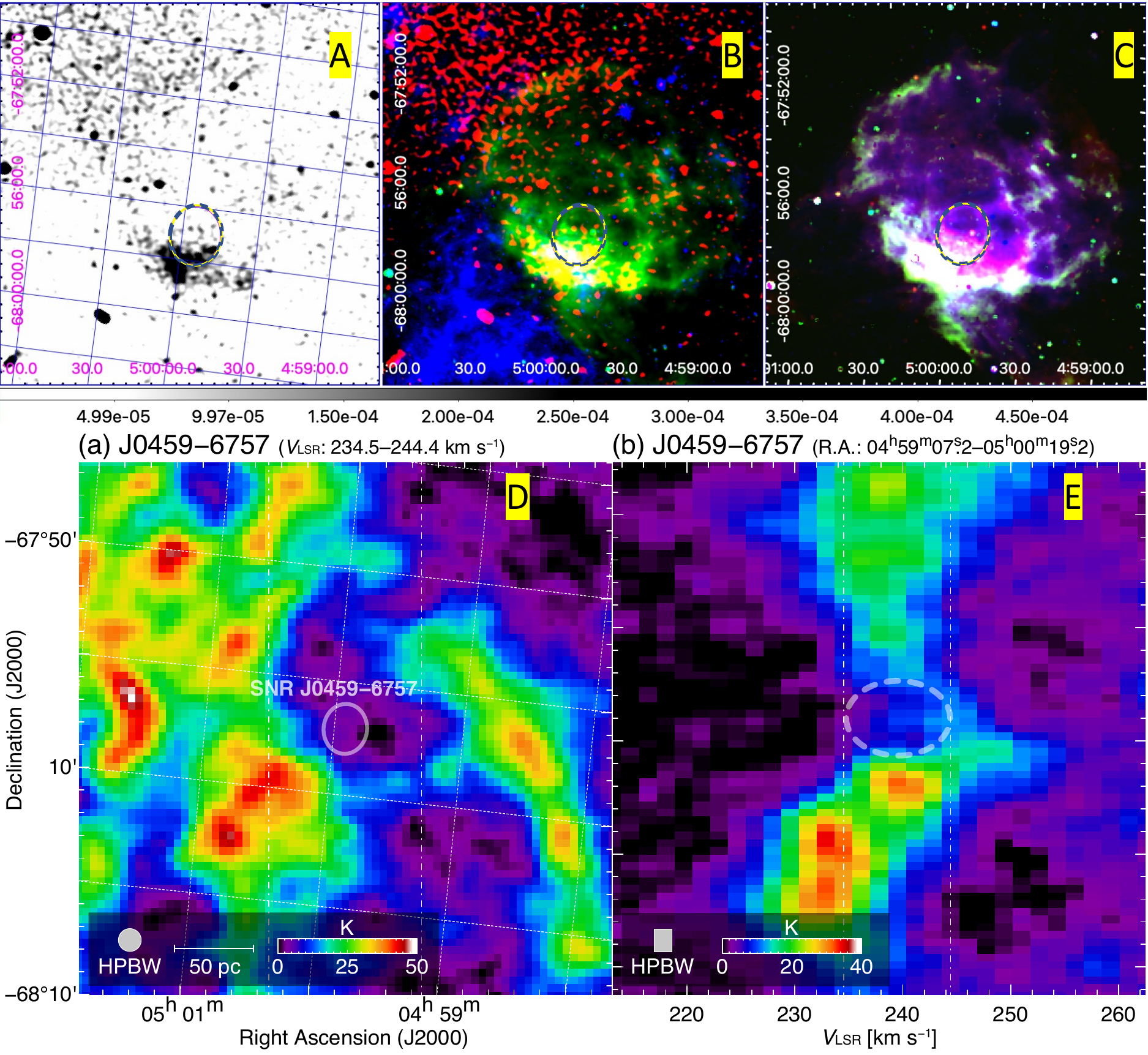}}
   \caption{J0459--6757: Same as Fig.~\ref{fig:0451-6906}. Gray scale at the bottom of panel A (\ac{ASKAP} image) is from 0 to 0.5~mJy~beam$^{-1}$.}
   \label{fig:0459-6757}
  \end{center}
\end{figure*}

This is the smallest \ac{SNR} candidate in our sample having a diameter of 30~pc embedded in the massive \HII\ region N\,16A. It has an elongated {half-ring} morphology, with the strongest emission seen toward the source's southern boundary where it appears to be interacting with a gas cloud  resulting in rim brightening (Figs.~\ref{fig:X} and~\ref{fig:0459-6757}). While the radio-to-\Ha\ ratio is only mildly enhanced, the \SII/\Ha\ ratio is not elevated indicating  the absence of radiative shocks. This is a very tentative candidate with more evidence shown by its optical properties. There are 18 OB star candidates located within, or in close proximity to this source but only one within the projected bounds. Both \HI\ map and $p-v$ diagram indicate the presence of a cavity, so we tentatively classify this object as \ac{SNR} candidate.

\subsection{J0504--6901}

\begin{figure*}
  \begin{center}
    \vspace{3mm}
    \resizebox{0.925\linewidth}{!}{\includegraphics[trim = 0 0 0 0,clip]{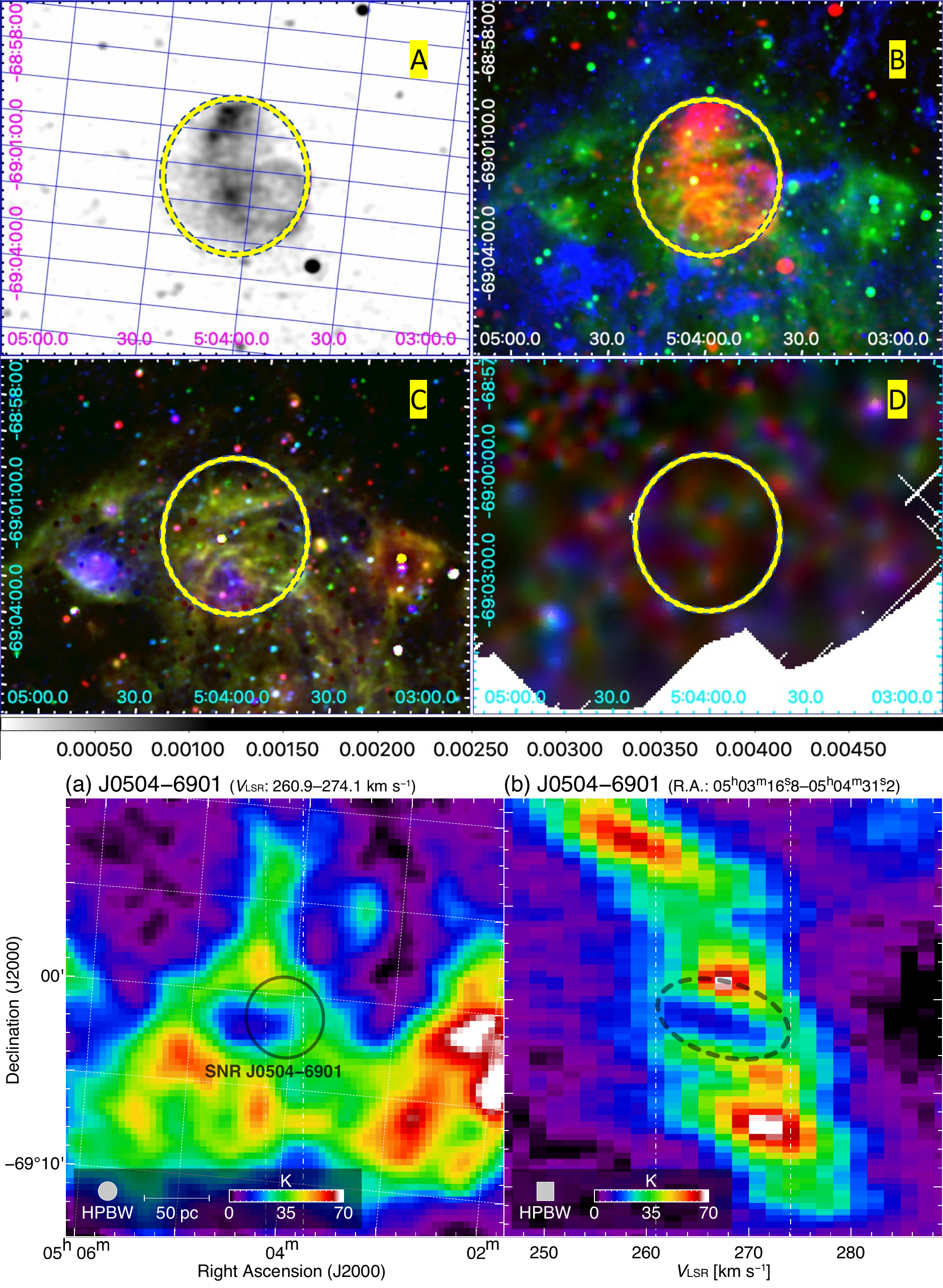}}
   \caption{J0504--6901: Same as Fig.~\ref{fig:0451-6951}. We note an image artefact in the X-ray (D) panel. Gray scale at the bottom of panel A (\ac{ASKAP} image) is from 0 to 5~mJy~beam$^{-1}$.}
   \label{fig:0504-6901}
  \end{center}
\end{figure*}

J0504--6901 is a potential remnant of an \ac{SN} that exhibits a complex and large ($D=61$\,pc) shell-morphology with a brightened region just off-centre and another in the northern part (Figs.~\ref{fig:X} and~\ref{fig:0504-6901}). It is placed in the larger region known as DEM\,L64 and originally was detected as a radio source (LMC~B0504--6906) in \citet{1995A&AS..111..311F,1996A&AS..120...77F,1998A&AS..130..421F} Parkes surveys. They found step spectral index of $\alpha$=--1.2$\pm$0.4 which supports a non-thermal radio spectrum, hence possible \ac{SNR} identification. However, at least one strong nearby radio source (J0503--6903) and a number of other fainter ones are within the large Parkes beams (ranging from 2.7~arcmin at 8640~MHz to 15.2~arcmin at 1400~MHz) and would significantly contribute the measured flux densities and therefore may not represent the true spectral index of this \ac{LMC} source alone. Very strong elevation of the radio-to-\Ha\ ratio warrants the \ac{SNR} candidate classification even though no clear X-ray (due to a low exposure time and a large off-axis angle) or optical boundaries could be drawn. There are 72 OB star candidates located within, or in close proximity to this source, with 10 located inside  the measured extent. Our \HI\ map and the $p-v$ diagram show a clear cavity in which this object may be expanding.

\subsection{MCSNR~J0522--6543}
\label{sec:0522notes}

\begin{figure*}
  \begin{center}
    \resizebox{0.9\linewidth}{!}{\includegraphics[trim = 0 0 0 0,clip]{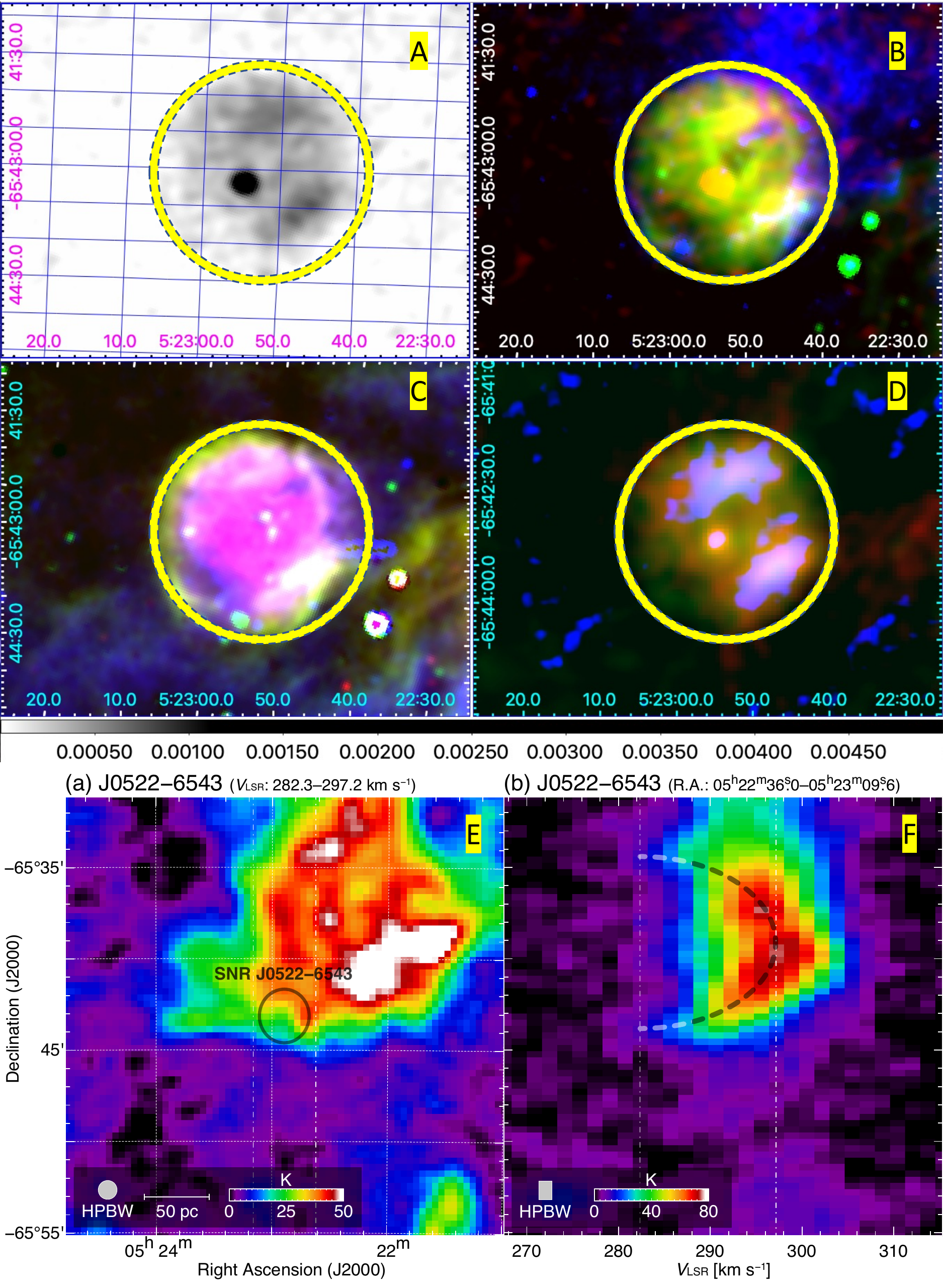}}
   \caption{MCSNR~J0522--6543:
    (Top) Left (A): New \ac{ASKAP} 888~MHz radio image at 13.9$\times$12.1~arcsec$^2$ resolution. Gray scale at the bottom of panel A (\ac{ASKAP} image) is from 0 to 5~mJy~beam$^{-1}$.
    Right (B): RGB image where \ac{ATCA} 2100~MHz radio image is in red, \Ha\ (green) and Spitzer LMC-SAGE at 8$\mu$m in blue. 
    (Middle) Left (C): \ac{MCELS} optical RGB image where \Ha\ (red), \SII\ (green), \& \OIII\ (blue). 
    Right (D): RGB image made from our new \ac{ATCA} observations at 2100~MHz (red), 5500~MHz (green) and 9000~MHz (blue). All images are smoothed to the resolution of 2100~MHz image (20.95$\times$16.60 arcsec$^2$).
    The blue-yellow ellipse indicates positions of here proposed \ac{SNR}.
   (Bottom) Left (E): Integrated intensity map of \HI\ obtained with \ac{ATCA} \& Parkes \citep{2003ApJS..148..473K}. 
   Right (F): Position--velocity diagram of H{\sc i}. 
   }
   \label{fig:0522-6543}
  \end{center}
\end{figure*}

\begin{figure*}
  \begin{center}
    \resizebox{1\linewidth}{!}{\includegraphics[trim = 0 0 0 0,clip]{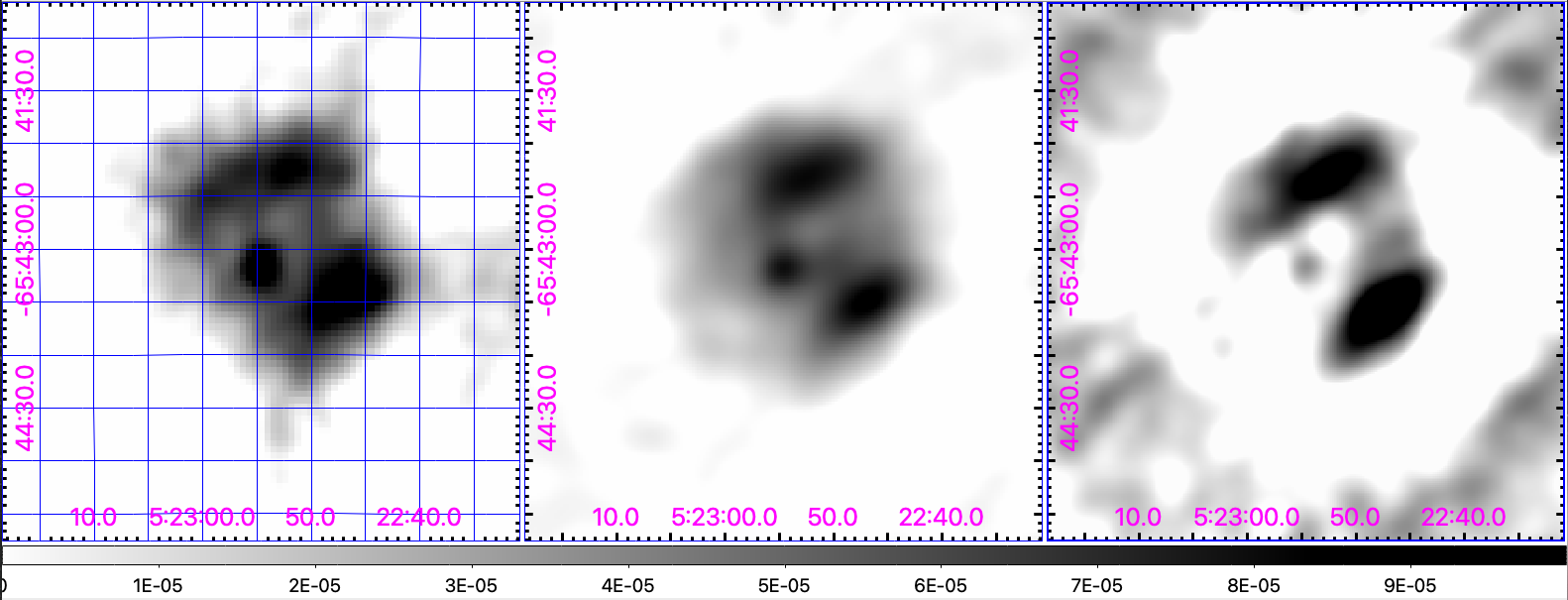}}
   \caption{
   \ac{ATCA} images of MCSNR~J0522--6543 at 2100~MHz (left), 5500~MHz (middle) and 9000~MHz (right). All images are smoothed to the resolution of 2100~MHz image (20.95$\times$16.60 arcsec$^2$). Gray scale at the bottom is from 0 to 0.1~mJy~beam$^{-1}$.
    }
   \label{fig:atca}
  \end{center}
\end{figure*}

This candidate has a shell morphology with ring brightened regions in the north-east and south-west (Figs.~\ref{fig:X}, \ref{fig:0522-6543} and \ref{fig:atca}). It also has a dominant bright central source. The intriguing central position of this point-like source suggests the possible presence of a \ac{PWN}, but we can not exclude a random background galaxy. This region was catalogued by \citet{1976MmRAS..81...89D} and given the reference DEM\,L155A. They described the source as a bright diffuse circular region $2.5\times2.5$~arcmin$^2$, which is fractionally smaller than our diameter measurement of $2.85\times2.63$~arcmin$^2$. The \ac{MCELS} image has an enhanced \SII/\Ha\ ratio of just slightly above 0.4 in the sources north-east region. There are 15~OB star candidates located within, or in close proximity to this source with 7 inside the projected boundary and four of these just off centre. Also, some 18~arcsec from the centre is the well studied young star cluster KMHK\,833 \citep{1990A&AS...84..527K}. MCSNR~J0522--6543 shows clear evidence of expanding inside a cavity or bubble-like structure in the \HI\ map and the $p-v$ diagram. These \HI\ clouds are most likely associated with the \ac{SNR} as they seems to be positioned along  the edge of the somewhat larger expanding \HI\ bubble (see Fig.~\ref{fig:0522-6543} panels D and E).

We measured flux densities (see Table~\ref{tab:J0522}) and obtained a spectral index of $\alpha$=--0.51$\pm$0.05 for MCSNR~J0522--6543 and for the central bright point source we estimate $\alpha$=--1.04$\pm$0.04. These results indicate the spectral index is typical of \ac{LMC} \acp{SNR} \citep{2017ApJS..230....2B} while the central bright source is most likely an unrelated background object (\ac{AGN} or radio quasar). However, as discussed above (Section~\ref{sec:results}), missing short spacing may influence our flux densities estimates. This is especially true at the higher frequencies (\ac{ATCA}; 5500 and 9000~MHz; Fig.~\ref{fig:atca}) where our flux density measurements could be underestimated. \citet{2022MNRAS.512..265F} suggest these could be as much as $\sim$15--20~per~cent (in flux density) which translates to a flattening of the spectral index by about 0.1 to $\alpha\sim$--0.41. Nonetheless, this would still allow the spectral index for MCSNR~J0522--6543 to remain in a range acceptable for \acp{SNR} classification.

\begin{table}
	\centering
	\caption{Radio flux density measurements for MCSNR~J0522--6543 and for the central point source J052254.7--654311.  
	}
	\label{tab:J0522}
	\begin{tabular}{lccccrccccc} 
		\hline
                            & MCSNR~J0522--6543     & J052254.7--654311   \\		
		\hline
	\noalign{\smallskip}
$S_{\rm 888\,MHz}$ (mJy)    & 33.1$\pm$0.7          & 3.52$\pm$0.11 \\
$S_{\rm 1384\,MHz}$ (mJy)   & 31.3$\pm$0.4          & 2.10$\pm$0.05 \\
$S_{\rm 2100\,MHz}$ (mJy)   & 19.8$\pm$0.4          & 1.47$\pm$0.05 \\
$S_{\rm 5500\,MHz}$ (mJy)   & 15.1$\pm$0.4          & 0.55$\pm$0.05 \\
$S_{\rm 9000\,MHz}$ (mJy)   & 10.2$\pm$0.4          & 0.29$\pm$0.05 \\
$\alpha\pm\Delta\alpha$     & --0.51$\pm$0.05       & --1.04$\pm$0.04\\
	\noalign{\smallskip}
		\hline
	\end{tabular}
\end{table} 

MCSNR~J0522--6543 satisfies two out of three criteria (optical \SII/\Ha$>$0.4 and radio spectral index $\alpha<$--0.5) to be a bona fide \ac{SNR}. Future deep X-ray observations of this region may reveal more about the true nature of this source.

\subsection{J0534--6720}

\begin{figure*}
  \begin{center}
    \resizebox{0.975\linewidth}{!}{\includegraphics[trim = 0 0 0 0,clip]{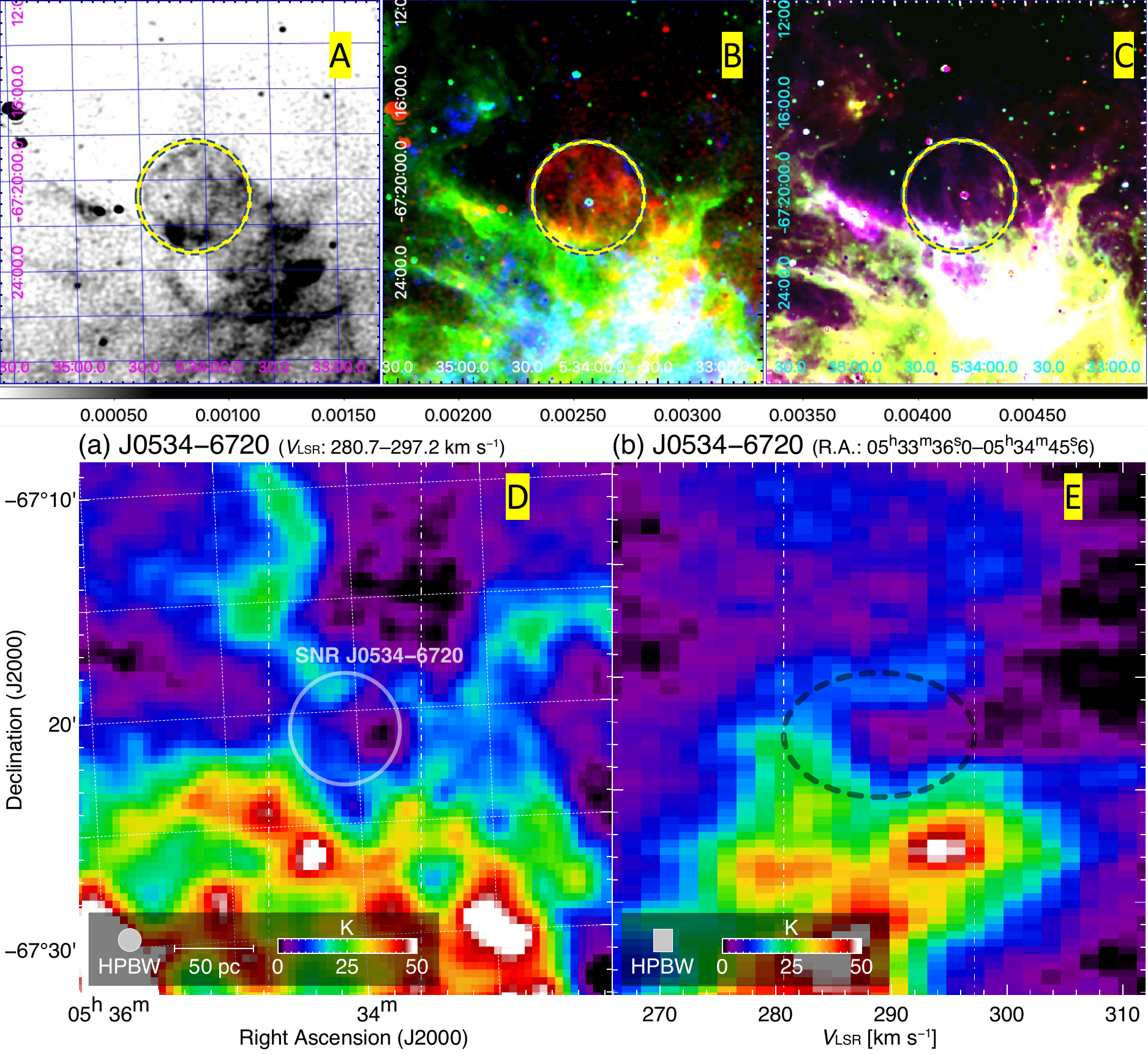}}
   \caption{J0534--6720: Same as Fig.~\ref{fig:0451-6906}. Gray scale at the bottom of panel A is from 0 to 5~mJy~beam$^{-1}$. The central star visible in optical images is a Galactic foreground star HD~269767 and is not related to the source.}
   \label{fig:0534-6720}
  \end{center}
\end{figure*}

This is the second largest candidate proposed in this study at $D=70$\,pc and appears with a complex shell morphology that includes areas of enhanced emission around the rim (Figs.~\ref{fig:X} and~\ref{fig:0534-6720}). The radio shell is most pronounced to the south. There is also noticeable radio emission towards the west and slightly beyond the presumed circular boundaries of this \ac{SNR} candidate that might be an integral part of the object. Its pear-like appearance is reminiscent of the Galactic \ac{SNR} Cygnus Loop, which might consist of two \acp{SNR} \citep{2002A&A...389L..61U}.
There are 45 OB star candidates located within 100~pc of the source, but only one appears within the bounds of the emission. While the enhanced radio-to-\Ha\ ratio shows a large ring-like shell, no real traces of radiative shocks are seen in any \ac{MCELS} band. But, the \HI\ map and the $p-v$ diagram show a clear cavity structure at this position which warrants an \ac{SNR} candidate classification. This object is therefore potentially an excellent example of an \ac{SNR} expanding inside the large scale cavity of a (super)bubble.

\subsection{J0534--6700}

\begin{figure*}
  \begin{center}
    \resizebox{0.975\linewidth}{!}{\includegraphics[trim = 0 0 0 0,clip]{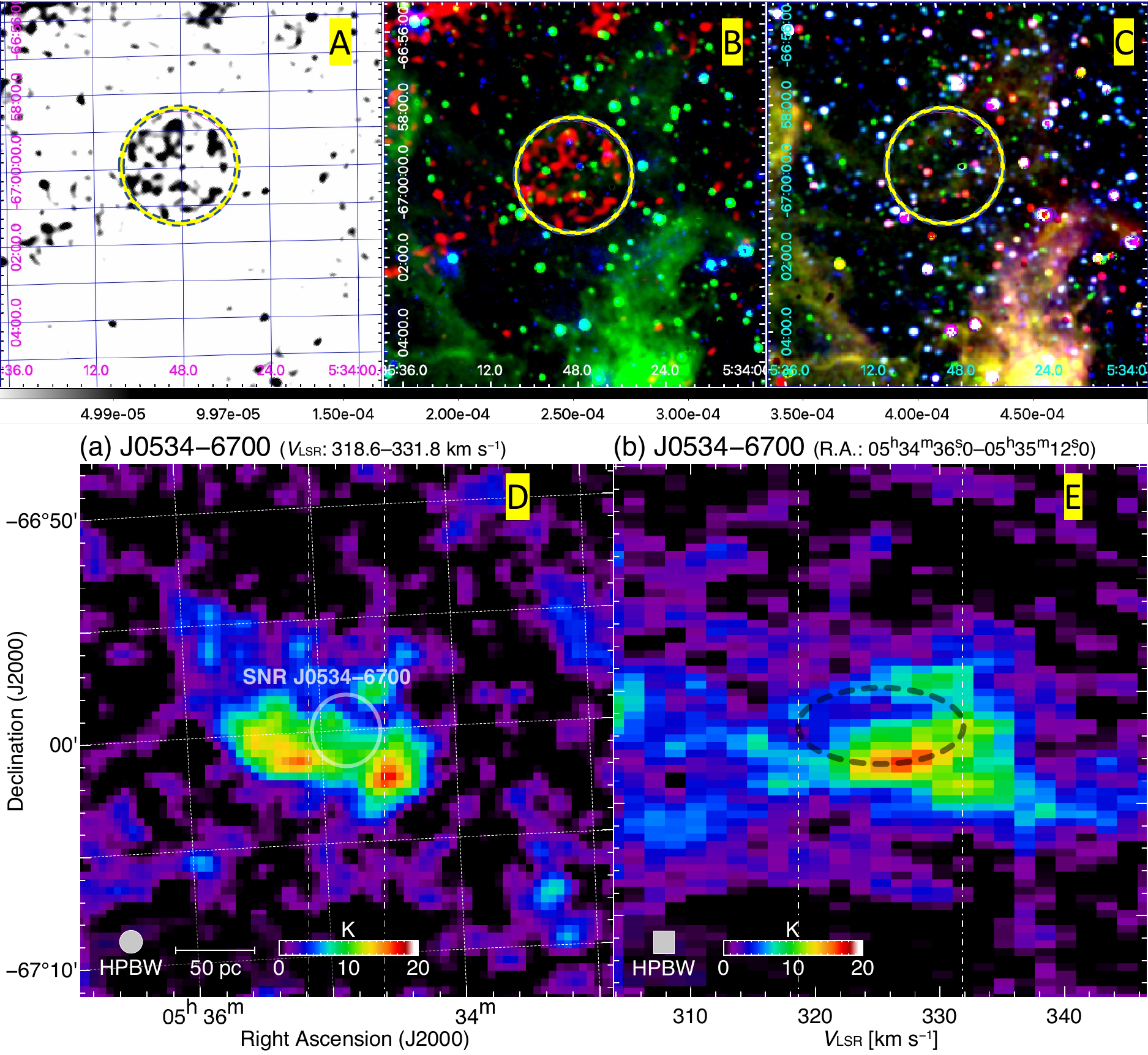}}
   \caption{J0534--6700: Same as Fig.~\ref{fig:0451-6906}. Gray scale at the bottom of panel A (\ac{ASKAP} image) is from 0 to 0.5~mJy~beam$^{-1}$.   }
   \label{fig:0534-6700}
  \end{center}
\end{figure*}

This faint and almost circular shell morphology source (Figs.~\ref{fig:X} and~\ref{fig:0534-6700}) is located at the eastern end of the large LMC4 supergiant shell \citep{1980MNRAS.192..365M}. There are 325 OB star candidates located within, or in close proximity to, this source. As for the number of other sources in our sample, we can see a clear cavity in our \HI\ and $p-v$ diagram. The absence of an optical detection is intriguing, but expected for this possible late evolutionary stage \ac{SNR}.

\subsection{J0542-6852}

\begin{figure*}
  \begin{center}
    \vspace{3mm}
    \resizebox{0.975\linewidth}{!}{\includegraphics[trim = 0 0 0 0,clip]{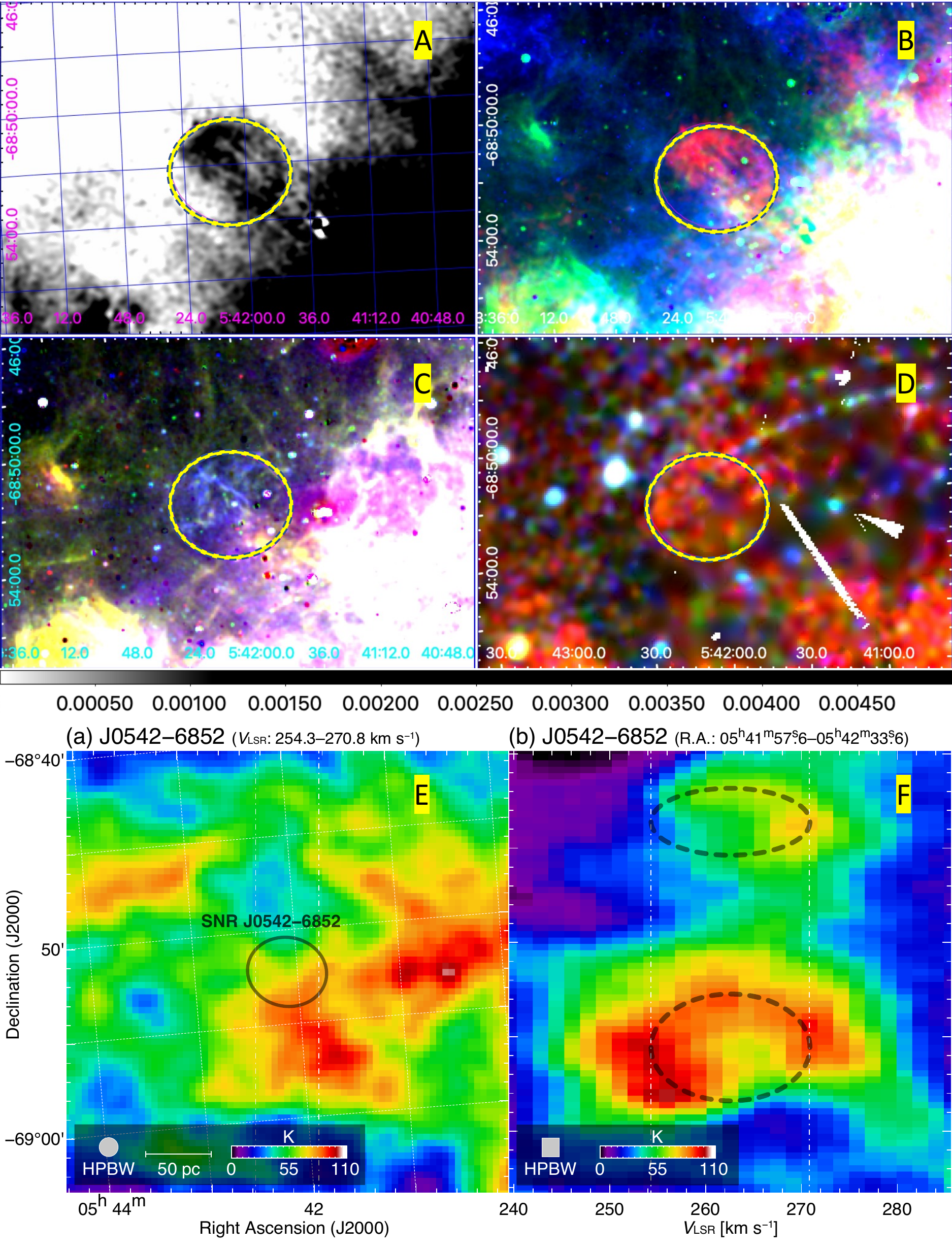}}
   \caption{J0542--6852: 
      Same as Fig.~\ref{fig:0451-6951}. Gray scale at the bottom of panel A (\ac{ASKAP} image) is from 0 to 5~mJy~beam$^{-1}$. We note an image artefact in the X-ray (D) panel.}
   \label{fig:0542-6852}
  \end{center}
\end{figure*}

The radio emission of this candidate appears with a faint and slightly elongated shell-like morphology (Figs.~\ref{fig:X} and~\ref{fig:0542-6852}). The remnant also appears to show faint X-ray emission in the soft and medium bands of \xmm\ survey images as well as some filaments associated with \OIII\ emission. There are 40 OB star candidates located within, or in close proximity to this source. J0542-6852 has possibly two counterparts of wind bubbles in the $p-v$ diagram, but these spatial extents in declination are slightly offset from the \ac{SNR} position. It is, therefore, possible that these two expanding bubbles are not from the \ac{SNR} candidate, but were caused by other events such as past activities of nearby OB associations.

\subsection{J0543--6928}

\begin{figure*}
  \begin{center}
    \vspace{3mm}
    \resizebox{0.925\linewidth}{!}{\includegraphics[trim = 0 0 0 0,clip]{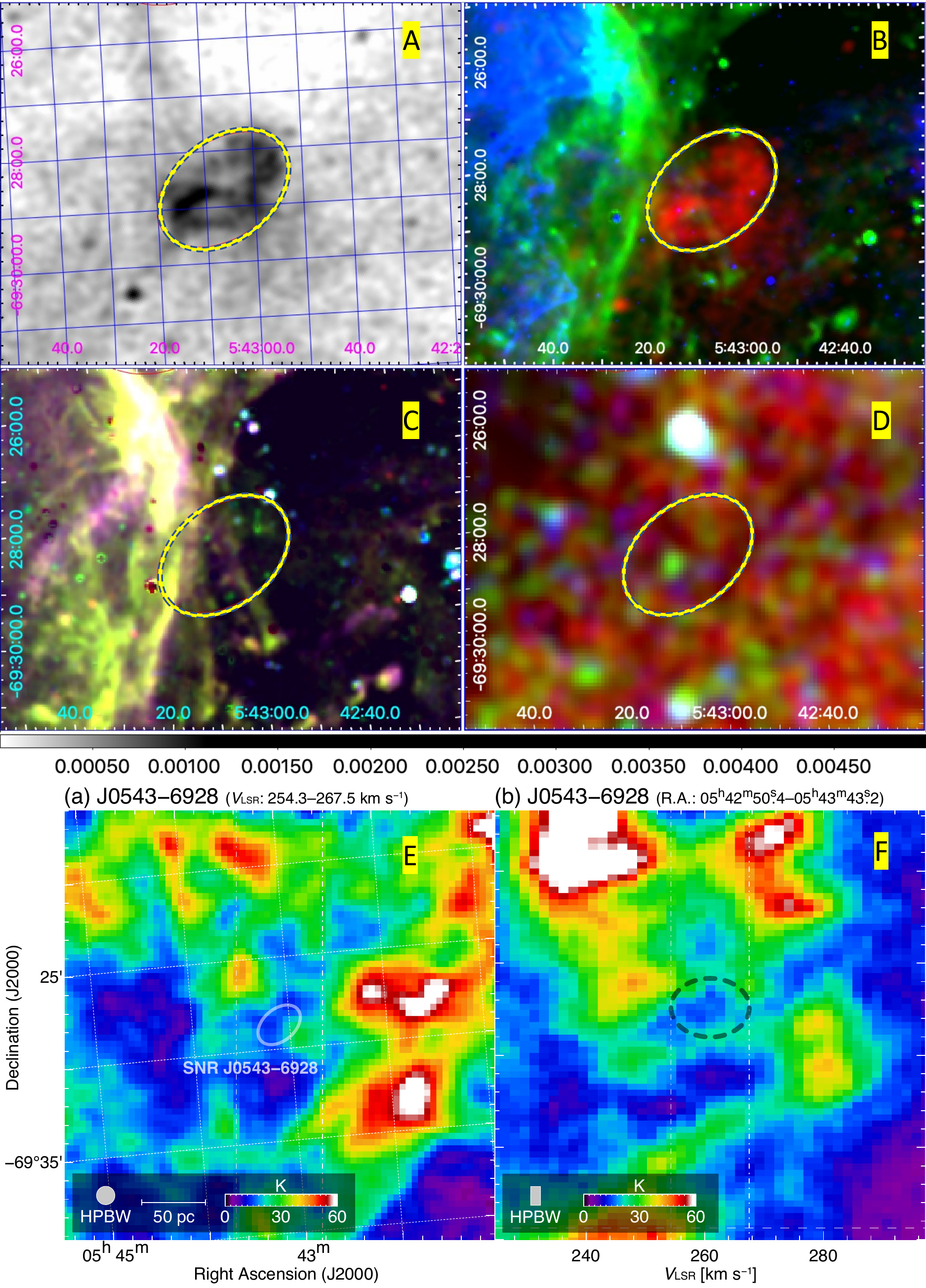}}
   \caption{J0543--6928: 
    Same as Fig.~\ref{fig:0451-6951}. Gray scale at the bottom of panel A (\ac{ASKAP} image) is from 0 to 5~mJy~beam$^{-1}$.
    }
   \label{fig:0543-6928}
  \end{center}
\end{figure*}

This source exhibits a complex shell morphology elongated in the NW-SE direction (Figs.~\ref{fig:X} and~\ref{fig:0543-6928}). It has a very strong enhancement in the radio-to-\Ha\ ratio image which suggests the structure is non-thermal. However, no counterpart is seen in the \ac{MCELS}, nor in our deep \xmm\ images. There are 34 OB star candidates located within, or in close proximity to, this source, though, only one of these lies in the bounds of the emission.

\subsection{J0543--6923}

\begin{figure*}
  \begin{center}
    \vspace{3mm}
    \resizebox{0.925\linewidth}{!}{\includegraphics[trim = 0 0 0 0,clip]{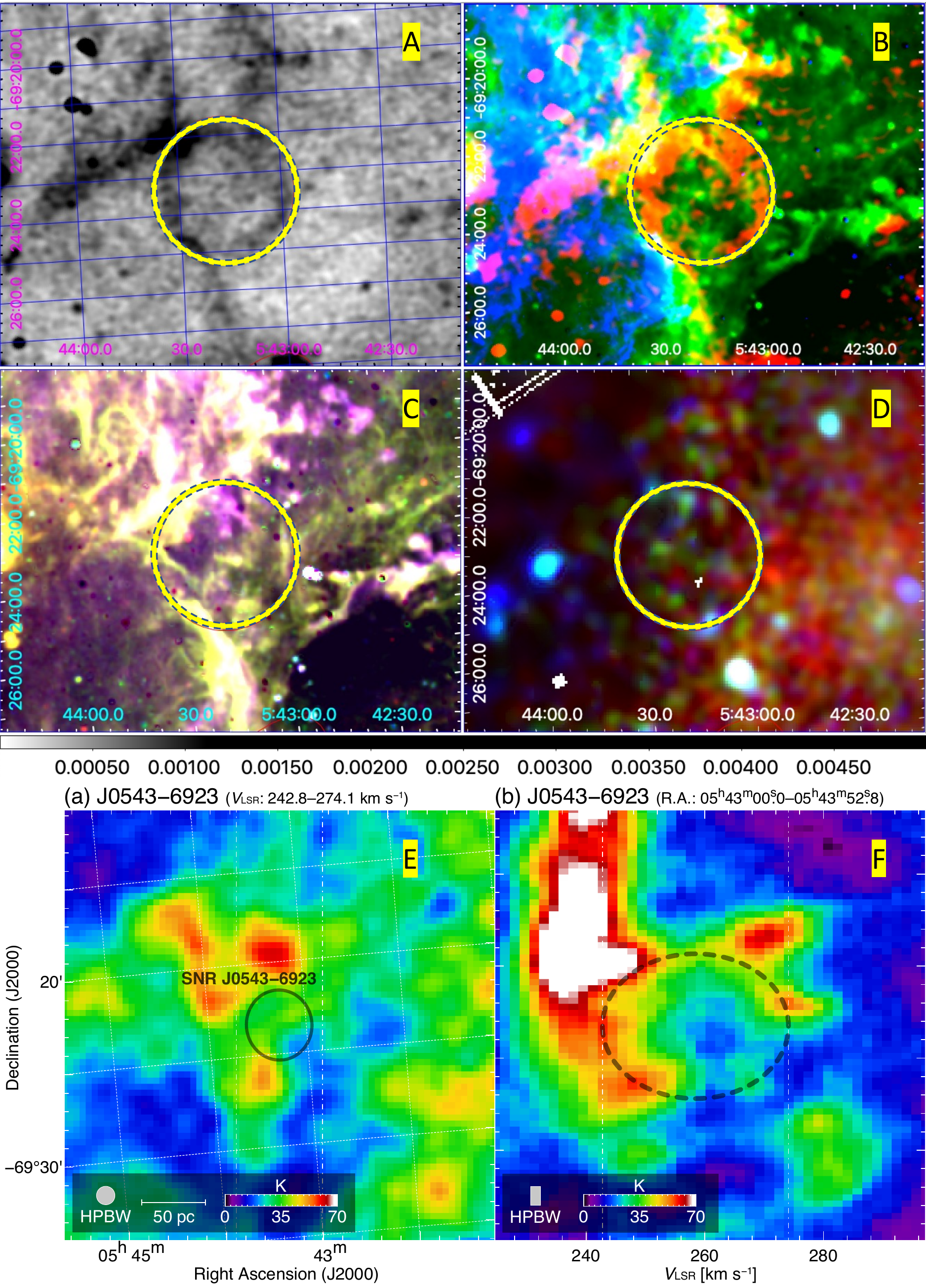}}
   \caption{J0543--6923: 
    Same as Fig.~\ref{fig:0451-6951}. Gray scale at the bottom of panel A (\ac{ASKAP} image) is from 0 to 5~mJy~beam$^{-1}$. We note an image artefact in the X-ray (D) panel.}
   \label{fig:0543-6923}
  \end{center}
\end{figure*}

Emission from this source follows a ring morphology (Figs.~\ref{fig:X} and~\ref{fig:0543-6923}). Brightening in our {radio/\Ha/IR image (panel B of Fig.~\ref{fig:0543-6923})}  can be seen in the north {and} east where the source appears to be colliding with emission from the \HI\ cloud {(panel E of Fig.~\ref{fig:0543-6923})}. Indeed, our $p-v$ diagram shows a clear cavity where this \ac{SNR} candidate may be expanding. There are 31 OB star candidates located within, or in close proximity to, this source. However, only one is within the bounds of the emission -- located at an area of brightening in the north-east. {We note a significant amount of possible non-thermal radio emission beyond the proposed boundaries (east) of this \ac{SNR} candidate. While we cannot exclude the association of these regions with the proposed remnant J0543--6923, we  suggest it could be  part of a larger and non-related (super)bubble structure. }

\subsection{J0543--6906}

\begin{figure*}
  \begin{center}
    \vspace{3mm}
    \resizebox{0.975\linewidth}{!}{\includegraphics[trim = 0 0 0 0,clip]{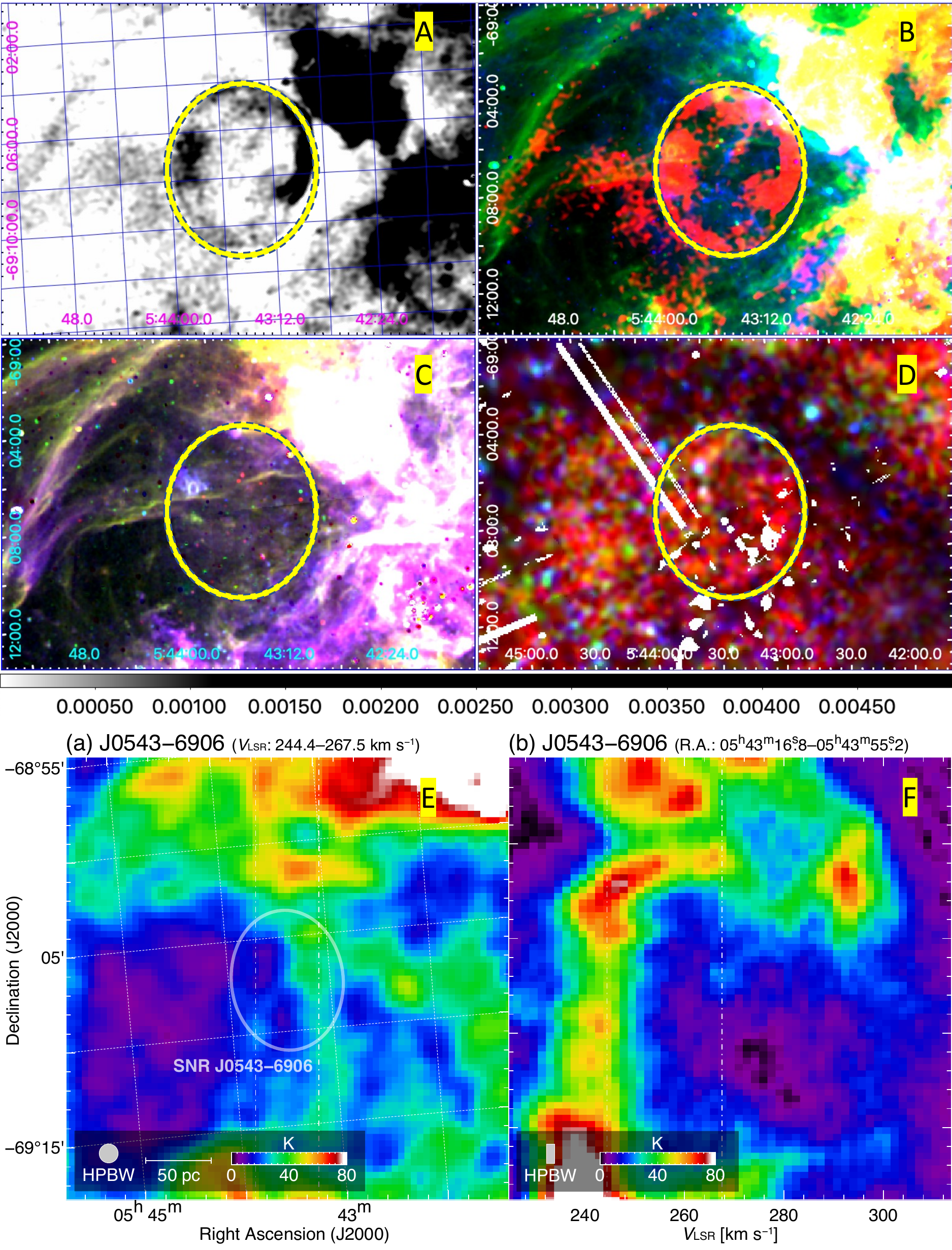}}
   \caption{J0543--6906: 
       Same as Fig.~\ref{fig:0451-6951}. Gray scale at the bottom of panel A (\ac{ASKAP} image) is from 0 to 5~mJy~beam$^{-1}$. We note  image artefacts in the optical (C) and X-ray (D) panels.}
   \label{fig:0543-6906}
  \end{center}
\end{figure*}

This source is the largest in our study ($D=96$\,pc) and exhibits a ring-like morphology with significant brightening on the western rim (Figs.~\ref{fig:X} and~\ref{fig:0543-6906}). While there are a large number of optical filaments across this region, we cannot associate any of them with this potentially largest \ac{SNR} {candidate in our sample}. Unfortunately, our \xmm\ images of this region are not sensitive enough to clearly confirm its nature. There are 50 OB star candidates located within, or in close proximity to this source, with eight inside the measured extent of emission. J0543--6906 is likely located in the large wind-(super)bubble that was formed by massive star cluster(s); its size is significantly larger than that of a single supernova.

\bsp	
\label{lastpage}
\end{document}